\documentclass[longauth]{aa}

\usepackage{natbib}
\bibpunct{(}{)}{;}{a}{}{,}

\usepackage{graphicx}
\usepackage{rotating}
\usepackage{multirow}
\usepackage{amssymb}
\usepackage{amsmath}
\usepackage{breakurl}
\usepackage{color}
\usepackage{threeparttable}
\usepackage{array,booktabs}

\def\apjl{ApJ}

\def\aap{A\&A}

\def\apjs{APJS}

\newcommand{\kms}{{\rm \, km~s}\ensuremath{^{-1}}}
\newcommand{\hinv}{\ensuremath{\, h^{-1}}}%
\newcommand{\msol}{\ensuremath{\, {\rm M}_\odot}}    
\newcommand{\msun}{\ensuremath{\, {\rm M}_\odot}} 
\newcommand{\kpc}{\ensuremath{\, {\rm kpc}}}         

\newcommand{\kev}{\ensuremath{\, {\rm keV}}}%

\newcommand{\sigmagal}{\ensuremath{\sigma_{\rm gal}}} 
\newcommand{\kt}{\ensuremath{{\rm k} T_X}}
\newcommand{\tthreeh}{\ensuremath{T_{\rm 300kpc}}}
\newcommand{\ktthreeh}{\ensuremath{{\rm k}T_{\rm 300kpc}}}
\newcommand{\vgal}{\ensuremath{v_{\rm gal}}}
\newcommand{\vmax}{\ensuremath{v_{\rm max}}}
\newcommand{\betaspec}{\ensuremath{\beta_{\rm spec}}}
\newcommand{\xamin}{{\sc Xamin} v3.3.2 }

\defcitealias{XXL-I:2016}{XXL~Paper~I}
\defcitealias{XXL-II:2016}{XXL~Paper~II}
\defcitealias{XXL-III:2016}{XXL~Paper~III}
\defcitealias{Lieu:2016}{XXL~Paper~IV}
\defcitealias{Guglielmo:inPrep}{XXL~Paper~XXII}
\defcitealias{Chiappetti:inPrep}{XXL~Paper~XXVII}
\defcitealias{Adami:inPrep}{XXL~Paper~XX}

\title{The XXL Survey\thanks{Based on observations obtained with XMM-Newton, an ESA science mission with instruments and contributions directly funded by ESA Member States and NASA.}}
\subtitle{XXIII. The Mass Scale of XXL Clusters from Ensemble Spectroscopy}

\begin{document}

   \author{Arya Farahi\inst{\ref{UMich}}
          \and Valentina Guglielmo \inst{\ref{INAPPadova},\ref{AixMarseille},\ref{UPadova}} 
          \and August E. Evrard\inst{\ref{UMich},\ref{UMichAstro}}
          \and Bianca M. Poggianti\inst{\ref{INAPPadova}}
          \and Christophe Adami\inst{\ref{AixMarseille}}
          \and Stefano Ettori\inst{\ref{INAFBologna}}
          \and Fabio Gastaldello\inst{\ref{IstitutodiAstrofisicaMilan}}
          \and Paul A. Giles\inst{\ref{Bristol}}
          \and Ben J. Maughan\inst{\ref{Bristol}}
          \and David Rapetti\inst{\ref{UColorado},\ref{NASAARC}}
          \and Mauro Sereno\inst{\ref{INAFBologna},\ref{UBologna}}
          \and Bruno Altieri\inst{\ref{ESA}}
          \and Ivan Baldry\inst{\ref{LiverpoolJMU}}
          \and Mark Birkinshaw\inst{\ref{Bristol}}
          \and Micol Bolzonella\inst{\ref{INAFBologna}}
          \and Angela Bongiorno\inst{\ref{INAFRome}}
          \and Michael J. I. Brown\inst{\ref{MonashU}} 
          \and Lucio Chiappetti\inst{\ref{IstitutodiAstrofisicaMilan}}
          \and Simon P. Driver\inst{\ref{ICRAR},\ref{UstAndrews}}
          \and Andrii Elyiv\inst{\ref{UBologna},\ref{ASUkraine}}
          \and Bianca Garilli\inst{\ref{IstitutodiAstrofisicaMilan}}
          \and Lo{\"i}c Guennou\inst{\ref{UKwaZuluNat}}
          \and Andrew Hopkins\inst{\ref{AAObservatory}}
          \and Angela Iovino\inst{\ref{INAFMilano}}
          \and Elias Koulouridis\inst{\ref{CEASaclay2},\ref{CEASaclay}}
          \and Jochen Liske\inst{\ref{UHamburg}}
          \and Sophie Maurogordato\inst{\ref{UNice}}
          \and Matthew Owers\inst{\ref{MacquarieU}}
          \and Florian Pacaud\inst{\ref{UBonn}}
          \and Marguerite Pierre\inst{\ref{CEASaclay2},\ref{CEASaclay}}
          \and Manolis Plionis\inst{\ref{AristotleU},\ref{Mexico},\ref{NOAthens}}
          \and Trevor Ponman\inst{\ref{UBirmingham}}
          \and Aaron Robotham\inst{\ref{ICRAR}}
          \and Tatyana Sadibekova\inst{\ref{CEASaclay2},\ref{CEASaclay},\ref{UlughBeg}}
          \and Marco Scodeggio\inst{\ref{IstitutodiAstrofisicaMilan}}
          \and Richard Tuffs\inst{\ref{MaxPlanck}}
          \and Ivan Valtchanov\inst{\ref{ESA}}
          }

   \institute{  
Department of Physics and Michigan Center for Theoretical Physics, University of Michigan, Ann Arbor, MI 48109, USA\label{UMich}
\email{aryaf@umich.edu}
\and
INAF - Osservatorio astronomico di Padova, Vicolo Osservatorio 5, IT-35122 Padova, Italy\label{INAPPadova}
\and
Aix Marseille Université, CNRS, LAM (Laboratoire
d’Astrophysique de Marseille) UMR 7326, F-13388, Marseille, France\label{AixMarseille}
\and
Department of Physics and Astronomy, University of Padova, Vicolo Osservatorio 3, IT-35122 Padova, Italy\label{UPadova}
\and
Department of Astronomy, University of Michigan, Ann Arbor, MI 48109, USA\label{UMichAstro} 
\and
INAF - Osservatorio Astronomico di Bologna, via Ranzani 1, IT-40127 Bologna, Italy\label{INAFBologna}
\and
Istituto di Astrofisica Spaziale e Fisica Cosmica Milano, Via Bassini 15, IT-20133 Milan, Italy\label{IstitutodiAstrofisicaMilan} 
\and
School of Physics, HH Wills Physics Laboratory, Tyndall Avenue, Bristol, BS8 1TL, UK\label{Bristol}
\and
Center for Astrophysics and Space Astronomy, Department of Astrophysical and Planetary Science, University of Colorado, Boulder, CO 80309, USA\label{UColorado}
\and
NASA Ames Research Center, Moffett Field, CA 94035, USA\label{NASAARC}
\and
Dipartimento di Fisica e Astronomia, Universit{\`a} di Bologna, viale Berti Pichat 6/2, IT-40127 Bologna, Italy\label{UBologna}
\and
Herschel Science Centre, European Space Astronomy Centre, ESA, ES-28691 Villanueva de la Ca{\~n}ada, Spain\label{ESA}
\and
Astrophysics Research Institute, Liverpool John Moores University, IC2, Liverpool Science Park, 146 Brownlow Hill, Liverpool L3 5RF, UK\label{LiverpoolJMU}
\and
INAF - Osservatorio Astronomico di Roma, via Frascati 33, IT-00078 Monte Porzio Catone (Rome), Italy\label{INAFRome}
\and
School of Physics, Monash University, Clayton, Victoria AU-3800, Australia\label{MonashU}
\and
International Centre for Radio Astronomy Research (ICRAR), The University of Western Australia, M468, 35 Stirling Highway, Crawley, WA AU-6009, Australia\label{ICRAR}
\and
SUPA, School of Physics and Astronomy, University of St Andrews, North Haugh, St Andrews KY16 9SS, UK\label{UstAndrews}
\and
Main Astronomical Observatory, Academy of Sciences of Ukraine, 27 Akademika Zabolotnoho St., UA-03680 Kyiv, Ukraine\label{ASUkraine}
\and
Max-Planck Institut f{\"u}er Kernphysik, Saupfercheckweg 1, DE-69117 Heidelberg, Germany\label{MaxPlanck}
\and
Astrophysics and Cosmology Research Unit, University of
KwaZulu-Natal, ZA-4041 Durban, South Africa\label{UKwaZuluNat}
\and
Australian Astronomical Observatory, PO-BOX 915, North Ryde, AU-1670, Australia\label{AAObservatory}
\and 
INAF, Osservatorio Astronomico di Brera, via Brera, 28, IT-20159 Milano, Italy\label{INAFMilano}
\and 
Universit{\"a}t Hamburg, Hamburger Sternwarte, Gojenbergsweg 112, DE-21029 Hamburg, Germany\label{UHamburg}
\and 
Laboratoire Lagrange, UMR 7293, Université de Nice Sophia Antipolis, CNRS, Observatoire de la C{\^o}te d’Azur, FR-06304 Nice, France\label{UNice}
\and 
Department of Physics and Astronomy, Macquarie University, NSW 2109, Australia and Australian Astronomical Observatory PO Box 915, North Ryde NSW AU-1670, Australia\label{MacquarieU}
\and 
Argelander Institut f{\"u}r Astronomie, Universit{\"a}t Bonn, Auf dem Huegel 71, DE-53121 Bonn, Germany\label{UBonn}
\and 
IRFU, CEA, Universit\'e Paris-Saclay, F-91191 Gif-sur-Yvette, France  \label{CEASaclay2}
\and
Universit\'e Paris Diderot, AIM, Sorbonne Paris Cit\'e, CEA, CNRS, F-91191
Gif-sur-Yvette, France\label{CEASaclay}
\and 
Aristotle University of Thessaloniki, Physics Department, GR-54124 Thessaloniki, Greece\label{AristotleU}
\and 
Instituto Nacional de Astrof{\'i}sica {\'O}ptica y Electr{\'o}nica, AP 51 y 216, 72000 Puebla, Mexico\label{Mexico}
\and 
IAASARS, National Observatory of Athens, GR-15236 Penteli, Greece\label{NOAthens}
\and 
School of Physics and Astronomy, University of Birmingham, Edgbaston, Birmingham B15 2TT, UK\label{UBirmingham}
\and 
Ulugh Beg Astronomical Institute of Uzbekistan Academy of Science, 33 Astronomicheskaya str., Tashkent, UZ-100052, Uzbekistan\label{UlughBeg}
}

   \date{Received September 15, 1996; accepted March 16, 1997}

\abstract
{An X-ray survey with the XMM-Newton telescope, XMM-XXL, has identified hundreds of galaxy groups and clusters in two 25 deg$^2$ fields. Combining spectroscopic and X-ray observations in one field, we determine how the kinetic energy of galaxies scales with hot gas temperature and also, by imposing prior constraints on the relative energies of galaxies and dark matter, infer a power-law scaling of total mass with temperature.}
{Our goals are: i) to determine parameters of the scaling between galaxy velocity dispersion and X-ray temperature, $\tthreeh$, for the halos hosting XXL-selected clusters, and; ii) to infer the log-mean scaling of total halo mass with temperature, $\langle \ln M_{200} \, | \, \tthreeh, z \rangle$.
}
{We apply an ensemble velocity likelihood to a sample of $> 1500$ spectroscopic redshifts within $132$ spectroscopically confirmed clusters with redshifts $z < 0.6$ to model, $\langle \ln\sigmagal\,|\,\tthreeh,z\rangle$, where $\sigmagal$ is the velocity dispersion of XXL cluster member galaxies and $\tthreeh$ is a 300 kpc aperture temperature. To infer total halo mass we use a precise virial relation for massive halos calibrated by N-body simulations along with a single degree of freedom summarizing galaxy velocity bias with respect to dark matter.  
}
{For the XXL-N cluster sample,  we find $\sigma_{\rm gal} \propto \tthreeh^{0.63\pm0.05}$, a slope significantly steeper than the self-similar expectation of $0.5$.  Assuming scale-independent galaxy velocity bias, we infer a mean logarithmic mass at a given X-ray temperature and redshift, $\langle \ln (E(z) M_{200}/10^{14} \msol)|\tthreeh,z\rangle=\pi_{T}+\alpha_{T}\ln\left(\tthreeh/T_p\right)+\beta_{T}\ln\left(E(z)/E(z_p)\right)$ using pivot values ${\rm k}T_{p}=2.2\,{\rm keV}$ and $z_p=0.25$, with normalization $\pi_{T}=0.45\pm0.24$ and slope $\alpha_{T}=1.89\pm 0.15$. We obtain only weak constraints on redshift evolution, $\beta_{T}=-1.29\pm 1.14$.}
{ The ratio of specific energies in hot gas and galaxies is scale dependent.  Ensemble spectroscopic analysis is a viable method to infer mean scaling relations, particularly for the numerous low mass systems with small numbers of spectroscopic members per system.  Galaxy velocity bias is the dominant systematic uncertainty in dynamical mass estimates.}


\date{Accepted. Received; in original form}

\titlerunning{The mass scale of XXL clusters}
\authorrunning{A. Farahi et al.}

\maketitle

\label{firstpage}

\section{Introduction}

The cosmic web of dark matter drives the gravitational potential wells in which baryonic matter is accelerated, shocked, stirred, and partially cooled into stars and galaxies.  Under gravity and shocks alone, the internal structure of collapsed halos is anticipated to be self-similar \citep{Bertschinger:1985}, meaning that the internal density and temperature profiles for hot gas maintain fixed forms in an appropriately scaled spatial radius. Integration of these forms enables straightforward calculation of global properties such as X-ray luminosity or temperature. The model specifies the slopes and evolution with redshift of key mass-observable relations \citep[MORs, see][]{Kaiser:1986}.

While astrophysical processes within halos, such as star formation and associated supernova and AGN feedback, are expected to drive  deviations from self-similarity, the observed population mean behavior of the most massive halos lie close to self-similar predictions \citep{Mantz:2016}.     

The idea that both galaxies and hot gas are in virial equilibrium within a common gravitational potential, originally proposed by \citet{Cavaliere:1976}, leads to the expectation that galaxy velocity dispersion scales as the square root of X-ray temperature,  $\sigmagal \propto T_X^{0.5}$.  This behavior reflects MOR scalings with total mass $M \propto T_X^{3/2}$ and $M \propto \sigmagal^3$ at fixed redshift.

For the most massive clusters in the sky, multiple surveys and follow-up observations are enabling individual halo masses to be estimated from gravitational lensing, hydrostatic, and dynamical methods \citep[see][for reviews]{Allen:2011,Kravtsov:2012}. 
These methods are subject to different sources of systematic uncertainty \citep[{\sl e.g.},][]{Meneghetti:2014}, and the samples to which they are applied may have additional systematic shifts, relative to a sample complete in halo mass, due to sample selection. The resulting biases pose limits on the accuracy of empirically derived MORs.   

Multiple, independent mass proxies allow for consistency tests that can expose and help mitigate systematic errors. We present here a virial analysis of 132 spectroscopically confirmed clusters identified in the XMM-XXL survey \citep[][hereafter XXL paper I]{XXL-I:2016}.  The method extends the stacked spectroscopic technique developed by \citet{Farahi:2016}, originally applied to optically selected clusters in SDSS \citep{Rykoff:2014}.

We focus first on the virial scaling of galaxy velocity dispersion with hot gas temperature, then infer how mean total mass scales with temperature using an additional degree of freedom that relates galaxy velocity dispersion to the underlying dark matter. This galaxy velocity bias is the largest source of uncertainty in our mass estimate.   

Early N-body simulations established virial scaling for purely dark matter halos \citep{Evrard:1989} and ensemble analysis of billion-particle and larger simulations provides a highly accurate calibration, with sub-percent error in the intercept of dark matter velocity dispersion at fixed halo mass \citep{Evrard:2008}. 

Inferring a virial, or dynamical, mass of an individual cluster requires a large number of spectroscopic members and a reliable interloper rejection algorithm \citep[{\sl e.g.},][]{Biviano:2006} such as that provided by the caustic technique \citep{Rines:2007, Rines:2010, Gifford:2013}. 
For large cluster samples emerging from surveys, a complementary approach to infer mean MOR scaling behavior is to employ ensemble population analysis, effectively stacking the local  velocities of galaxies in multiple clusters to extract a mean velocity dispersion signal \citep{Farahi:2016}. 

Here we employ a large collection of galaxy spectroscopic redshifts assembled from multiple sources for groups and clusters identified in the North field of the XMM-XXL survey.  The 132 systems span X-ray temperatures $\ktthreeh \in [0.48-6.03] \kev$, and redshift $z \in [0.03-0.6]$, and the spectroscopic sources include GAMA, SDSS-DR10, VIPERS, and VVDS Deep and Ultra Deep surveys. 

The mass-temperature scaling has been studied extensively \citep[{\sl e.g.},][]{Xue:2000,Ortiz-Gil:2004,Arnaud:2005,Vikhlinin:2006,Kettula:2015,Mantz:2016wtgV,Lieu:2016}. Observational relations generally steepen from close to the self-similar  for hot systems to a slope of $\sim 1.6-1.7$ once cooler systems $(\ktthreeh \lesssim 3 \kev)$ are included \citep{Arnaud:2005, Lieu:2016}. 
More than half of the clusters in this work will be systems with $\ktthreeh \lesssim 3 \kev$, which allows us to test deviation from the self-similar model, with yet another mass calibration technique.

As part of the first series of XXL papers, \citep[][hereafter XXL paper IV]{Lieu:2016} estimates the mass-temperature scaling relation of X-ray bright systems using weak-lensing mass measurements from the Canada-France-Hawaii Telescope Lensing Survey (CFHTLenS) shear catalog \citep{Heymans:2012, Erben:2013}.
The work presented here is complementary to that study where it provides a mean dynamical mass as a function of X-ray temperature. The X-ray sample differs from that used by \citetalias{Lieu:2016}, but the pipeline for deriving X-ray properties from the XMM data is identical. 

We describe the sample, data, and selection criteria in \S\ref{sec:data}.
The likelihood model used to constrain the galaxy velocity dispersion scaling with temperature is described in \S\ref{sec:like-model}. In \S\ref{sec:results}, we present  results for this relation, followed by a discussion of a range of systematic uncertainties and sensitivity analysis in \S\ref{sec:sensitivity-analysis}.
A key result of this work, the dynamical mass-temperature relation, is presented in \S\ref{sec:massCalibration}.
Finally we conclude in \S\ref{sec:conclusion} 

Throughout we assume WMAP9 consistent cosmology
with $\Omega_{m} = 0.28$, $\Omega_{DE} = 0.72$, and local Hubble constant $h = H_0/100$~km~s$^{-1}$~Mpc$^{-1}=0.7$.
Unless otherwise noted, our convention for the mass of
a halo is $M_{200}$, the mass contained within a spherical region encompassing a mean density equal to $200$ times the critical density of the Universe, $\rho_c(z)$. 
Similarly, $r_{\Delta}$ is defined as the radius of the sphere inside which the mean density is a factor $\Delta$ times the critical density of the universe at that redshift, and $M_{\Delta}$ is the total mass within that radius.

\section{Cluster and Spectroscopic Sample}\label{sec:data}

\begin{figure*}
\begin{center}
\includegraphics[scale=0.4]{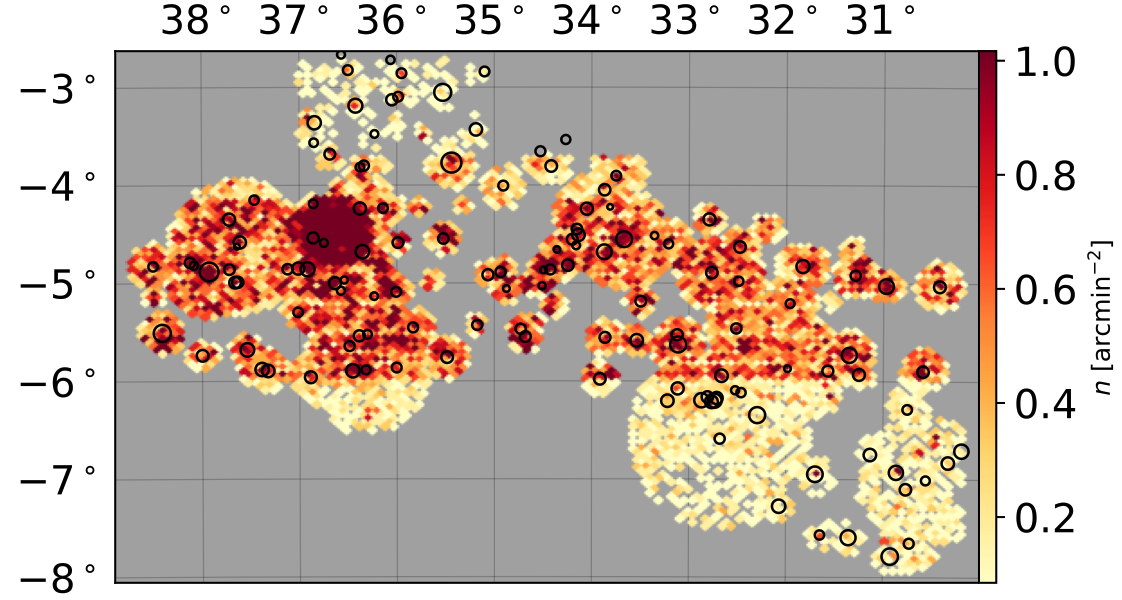}
\caption{Spatial distribution of galaxies and clusters in the XXL North field used in this work. Black circles show cluster centers with $z \leq 0.6$ with area proportional to temperature. The heat map shows the sky surface density of spectroscopic galaxies lying within a projected aperture of $3 r_{500}$  around cluster centers.  
}
\label{fig:RA_DEC_sample}
\end{center}
\end{figure*}

The XXL survey consists of tiled 10~ks (or longer) exposures across two fields of roughly 25 deg$^2$ each. The observing strategy and science goals of the survey are described in \citetalias{XXL-I:2016} while source selection and a resultant brightest 100 cluster sample are published in \citet[][hereafter XXL Paper II]{XXL-II:2016}.
The X-ray images were processed with the \xamin pipeline \citep{Pacaud:2006}, which produces lists of detections of varying quality. The overall catalog with point sources will be available in computer readable form via the XXL Master Catalog browser \url{http://cosmosdb.iasf-milano.inaf.it/XXL} and at the Centre de Donn\'ees astronomiques de Strasbourg (CDS)\footnote{\url{http://cdsweb.u-strasbg.fr}} \citep[][hereafter XXL Paper XXVIII]{Chiappetti:inPrep}, while cluster candidates are grouped by detection classes (C1, C2, C3) and hosted in the same places as catalog XXL-365-GC \citep[][hereafter XXL Paper XX]{Adami:inPrep}.
 The 2016 series of XXL papers, including \citepalias{XXL-II:2016}, pertained to the brightest 100 clusters and 1,000 AGN, while for the second series, including this paper, we are publishing much deeper samples: 365 clusters and 20,000 AGN, with slightly revised cluster properties and scaling relations.

Of the XXL cluster sample 46\% are classified as high-quality (C1) detections, 43\% are intermediate quality (C2) and the remaining 11\% are marginal quality (C3) sources. We discard C3 sources in this work as they do not have reliable luminosity and temperature measurements. The subject of this work is a subset in the XXL-N area, with spectroscopically confirmed redshifts and with redshifts $z < 0.6$, generating a sample of 132 systems. A detailed discussion of the sample selection is provided by \citetalias{Adami:inPrep} and \citet[][hereafter XXL Paper XXII]{Guglielmo:inPrep}. The cluster optical and X-ray images can be found in the XXL cluster database: \url{http://xmm-lss.in2p3.fr:8080/xxldb}.

The sky distribution of the systems used in this work is shown in Figure \ref{fig:RA_DEC_sample}.  X-ray extended sources are shown as black circles and the color map shows the sky surface density of spectroscopic galaxies lying in an aperture of radius $r \leq 3 r_{500}$ with respect to their centers. The $r_{500}$ estimates are determined from weak lensing mass estimates presented in \citetalias{Lieu:2016}.  We next provide additional details of the group/cluster and galaxy spectroscopic samples.

\subsection{X-ray Temperatures}\label{sec:temperatures}

Of the 132 spectroscopically confirmed C1 and C2 clusters with $z<0.6$, X-ray temperatures are available for 106, 81 C1 and 25 C2 clusters.
All are C1 clusters and most but not all are included in the XXL 100 brightest sample of \citetalias{XXL-II:2016}.  The temperature determination, described in detail by \citet[][hereafter XXL Paper III]{XXL-III:2016}, outputs the temperature measured within a physical 300 kpc aperture for sufficiently high signal-to-noise systems.

After detection by \xamin - a detection pipeline piloted by the 
XMM-LSS project \citep{Pacaud:2006} - as an extended X-ray source, a background subtracted radial profile is extracted in the $[0.5-2]~\kev$ band.  The detection radius is defined as that at which the source is detected at $5\sigma$ above the background.  A spectrum is then fit from a circular aperture of radius of 300 kpc centered on the X-ray centroid, using a minimum of 5 counts per energy bin, resulting in a temperature measurement we refer to as $\tthreeh$.  Cluster spectral fits were performed in the $0.4-7.0$ keV band with an absorbed APEC model with the absorbing column fixed at the Galactic value, and a fixed metal abundance of $Z=0.3Z_{\odot}$.
 For more detail on the data processing, we refer the reader to \citet{XXL-II:2016}. 
Note that the measured X-ray temperatures are non-core excised owing to the limited angular resolution of XMM-Newton and the modest signal-to-noise of most detections.  These temperatures are taken from \citetalias{Adami:inPrep}.

For the systems that lack direct temperature estimates, we 
estimate temperatures from X-ray luminosities using published XXL scaling relations as follows. First, background-corrected XMM count-rates within 300 kpc from the cluster center in the $[0.5-2]~\kev$ band are extracted.  This forms the basis of a first luminosity estimate, the starting point for an iterative scheme that uses the $L-T$ scaling relation from \citetalias{Adami:inPrep} and the $T-M_{500}$ relation from \citetalias{Lieu:2016}.  
The process assumes isothermal $\beta$-model emission with parameters $(r_c, \beta) = (0.15 r_{500}, 2/3)$, and iterations continue until convergence.  This method outputs temperature, mass, and $r_{500}$ estimates. Details of the steps above are described and reported in \citetalias{Adami:inPrep}.  

To check the internal consistency of the derived X-ray temperature, \citetalias{Adami:inPrep} performs a comparison of $\tthreeh$ derived using the above approach with direct temperature measurements for a subset of systems, finding good agreement. 
 Below, we show that the velocity dispersion scaling parameters using the subset of systems with directly measured temperatures are consistent with those of the full cluster sample.

\begin{figure}
 \centering
 \includegraphics[width=0.95\linewidth]{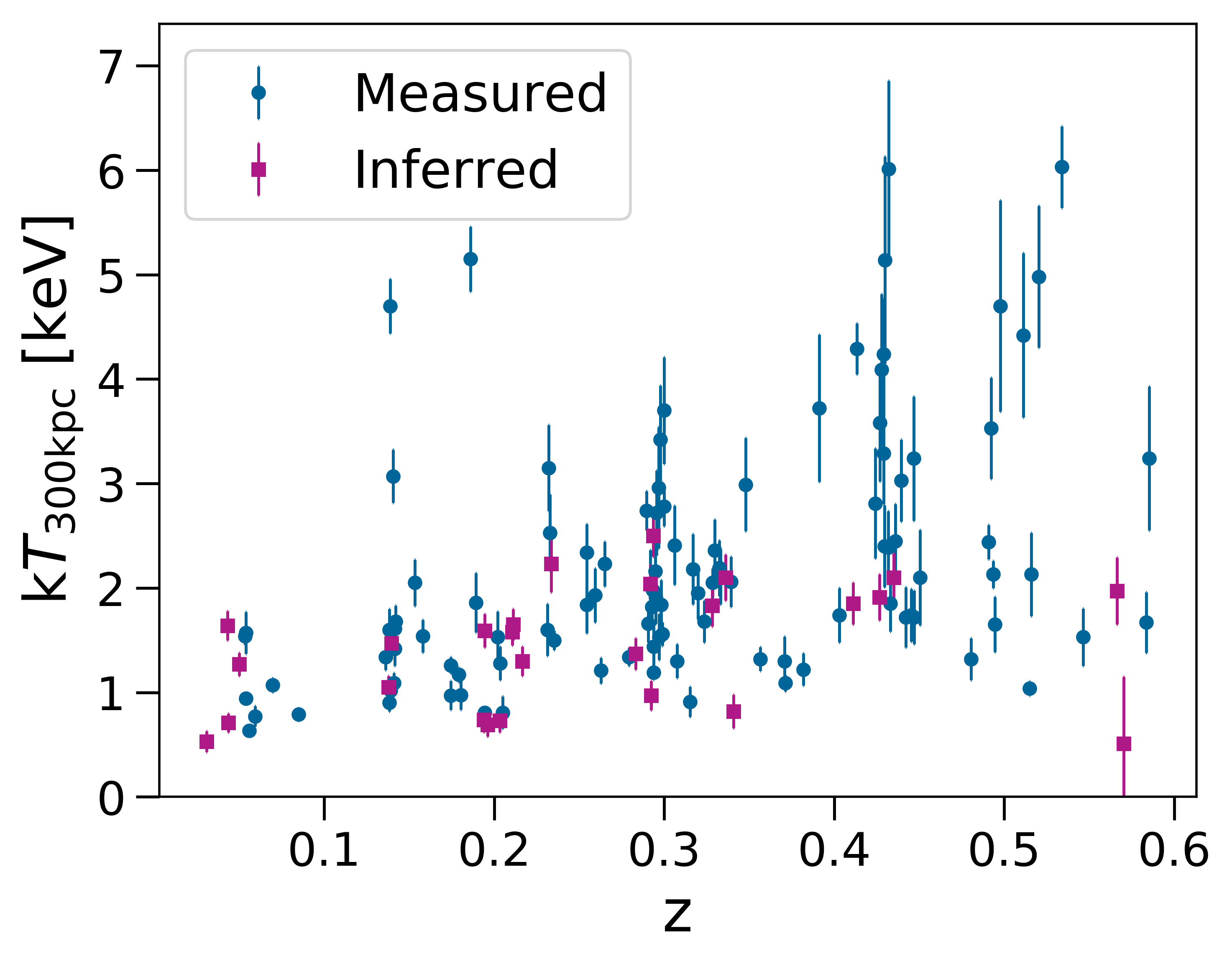}
 \caption{Temperature vs. redshift of the full 132 XXL-N cluster sample. Blue circles are clusters with measured temperature and magenta squares show clusters with inferred temperature. }
 \label{fig:sample}
\end{figure}

Figure~\ref{fig:sample} shows redshifts and temperatures of the XXL-N clusters.  At a given redshift, higher mass systems that are both brighter and hotter tend to have direct temperature measurements.  As explained in \S \ref{sec:signal}, the sample size shrinks, by roughly 3\% (4 clusters), after we apply velocity and aperture cuts discussed below.

\subsection{Spectroscopic Sample}

Concerning the spectroscopic database of galaxies, reduced spectra from several public surveys are combined with XXL dedicated observing runs to create a large, heterogeneous collection of redshifts.  The surveys and observing programs, listed in Table 2 in \citetalias{Guglielmo:inPrep}, include GAMA \cite[45\%,][]{Hopkins:2013,Liske:2015}, SDSS-DR10 \cite[5\%][]{Ahn:2014}, VIPERS \cite[32\%][]{Guzzo:2014}, VVDS Deep and Ultra Deep \cite[9\%][]{LeFevre:2005,LeFevre:2015}. The remaining 9\% are obtained mainly by ESO Large Program + WHT XXL dedicated observational campaigns which are individually contributing less than $2\%$. The typical error in redshift for galaxies is $\sim 0.00041 (1+z)$, equivalent to $120 (1+z)~{\rm km/s}$. The full list of spectroscopic catalogs are listed in \citetalias{Guglielmo:inPrep}. Note that the spectroscopic sample adopted in this work is a subset of the spectroscopic sample of \citetalias{Guglielmo:inPrep}.

Given that the catalog sources overlap in the sky, a non-negligible number of objects are observed by more than one project. The cleaning of catalog duplicates follows the selection criteria designed to identify the best spectrum in the final catalog, as described by \citetalias{Guglielmo:inPrep}. The selection procedure is based on two sets of priorities, the first regarding source origin and then the second regarding the reliability flag attributed to the redshift estimate. 

The full sample contains 120506 galaxies in the North XXL region, 63681 of which are at $z \leq 0.6$. 
For our default analysis, we employ a sub-sample comprised of those galaxies lying within a projected distance of $r_{500}$ from the centers of the clusters, shown in Figure~\ref{fig:RA_DEC_sample}, yielding 7751 galaxies. 

The spectroscopic information for these galaxies, as well as for spectroscopically confirmed groups/clusters, is hosted in the CeSAM (Centre de donn\'eeS Astrophysiques de Marseille) database in Marseille (CeSAM-DR2), publicly available at \url{http://www.lam.fr/cesam/}.

\subsection{Spectroscopic Redshifts of XXL-selected Clusters}

All C1 and C2 candidate clusters identified within the XXL survey are followed up for spectroscopic redshifts using an iterative semi-automatic process similar to that used for the XMM-LSS survey \citep{Adami:2011}.

First, spectroscopic redshifts from public and private sources lying within the X-ray contours are selected. 
These are sorted to identify significant (more than 3 galaxies) concentrations, including a preliminary ``cluster population'' based on projected separation from the X-ray centroid.
For the large majority of cases, a single concentration appears, allowing for relatively unambiguous redshift determination.

A preliminary measure of the cluster redshift is the \emph{mean} value of the redshift of the preliminary ``cluster population''. From this redshift, a physical region of $500~\kpc$ radius is defined, and all galaxies within this radius were selected as cluster members. This procedure is iterated with all available redshifts within a $500~\kpc$ physical radius to get the final mean cluster redshift.  However, for ambiguous cases where there are not more than three galaxies with spectroscopic redshifts, the redshift is measured by looking for the putative brightest cluster galaxy (BCG) in the i-band located close to the X-ray centroid (see \citetalias{Adami:inPrep} for a detailed discussion). 

The cluster center is defined by the peak in the detected X-ray emission.  Because X-ray emission is continuous and the gas traces the gravitational potential, we expect fewer mis-centered clusters (mis-centered with respect to the dark matter potential minimum) compared to photometrically-defined samples \citep{Rykoff:2012}.  We defer a detailed treatment of cluster mis-centering to future work.

\subsection{Galaxy-cluster Velocities}

Given the redshift, $z_c$, of each XXL-N group or cluster, we measure the rest-frame relative velocity of each galaxy within the target field of that cluster, 
\begin{equation}\label{eq:Velocity}
    \vgal \ = \ c \ \left( \frac{z_g - z_c}{1+z_c} \right),
\end{equation}
where $c$ is the speed of light and $z_g$ is the redshift of the galaxy. 

In this paper the original spectroscopic galaxy selection for each cluster is defined only by sky location, not cluster redshift.  Therefore, each cluster field contains a mix of galaxies residing within and outside the cluster environment. We describe below the probabilistic method originally applied to SDSS redMaPPer systems by \citet{Rozo:2015}, which involves a two-stage approach to handling foreground and background galaxies.

\subsection{Signal Component and Final Cluster Sample} \label{sec:signal}

The model framework, wherein observable properties scale with halo mass as power laws with some intrinsic covariance, motivates the modeling process.  For systems with a given temperature, $\tthreeh$, and redshift, we expect a log-normal distribution of halo mass with some intrinsic ($10-20\%$) scatter \citep{Lebrun:2016}.  The galaxy velocities internal to these halos are assumed to follow a Gaussian distribution with a dispersion that increases with halo mass. Because the intrinsic scatter of these relations is not very large, the expected distribution of galaxy velocities, $\vgal$, at fixed $\tthreeh$ and $z$ will also be close to Gaussian \citep[see][for a specific model applied to galaxy richness instead of temperature]{Becker:2007}.  This collective component is the fundamental signal we seek to model and extract from the data.  

\begin{figure}
 \centering
 \includegraphics[width=0.95\linewidth]{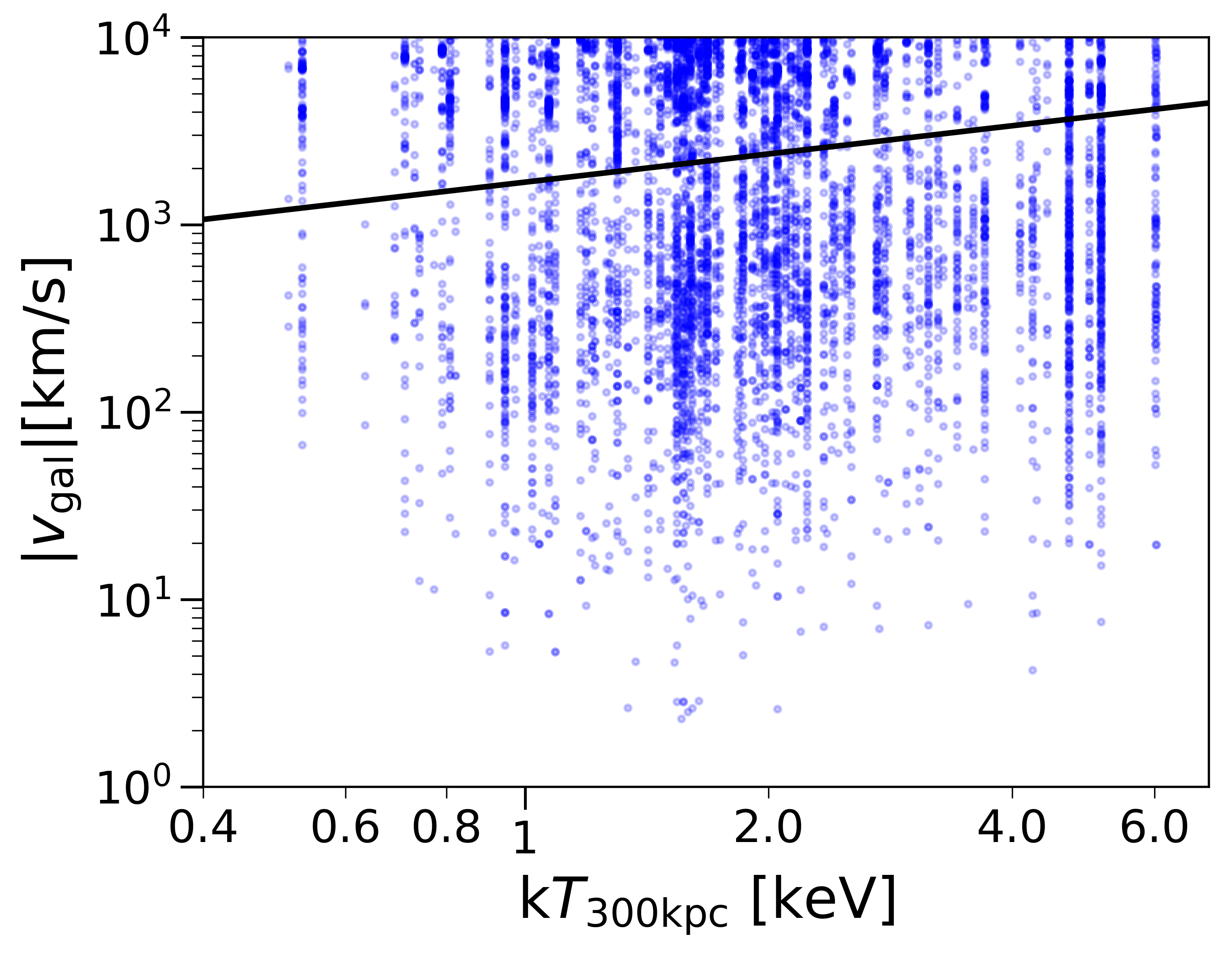}
 \caption{Magnitude of the rest-frame velocity of cluster galaxies, equation~(\ref{eq:Velocity}), as a function of cluster temperature.  Each dot is one galaxy, and some galaxies appear in the fields of multiple clusters.  The black line shows the cut, equation~(\ref{eq:Vmax}), that separates the lower signal population from a projected background. Points above the black line are disregarded in our analysis.}
 \label{fig:vel_structure}
\end{figure}

The first stage of the process removes projected interlopers with large $\vgal$ offsets, much larger than those expected from the underlying Gaussian model.  The threshold value, $\vmax(\tthreeh)$, is set empirically by examination of the absolute magnitude of the line-of-sight galaxy velocities as a function of cluster temperature, given in Figure~\ref{fig:vel_structure}.  Similar to the analysis of \citet{Farahi:2016}, where redMaPPer optical richness plays the role of $\tthreeh$, two populations emerge: a signal component at low velocities and a projected population offset to higher velocities. 

Based on the structure of Figure~\ref{fig:vel_structure}, we define a maximum, rest-frame galaxy velocity for the signal region of 
\begin{equation} \label{eq:Vmax}
    \vmax(\tthreeh) \ = \ 2500 ~\left(\frac{\ktthreeh}{2.2 \kev}\right)^{0.5} \kms.
\end{equation}
Applying this cut along with the radial cut, $r \le r_{500}$, eliminates 4 clusters from the sample because no galaxies satisfy these cuts. The final cluster sample involves 1592 galaxies across 128 clusters, 103 of which have directly measured temperatures. 

\begin{figure}
 \centering
 \includegraphics[width=0.95\linewidth]{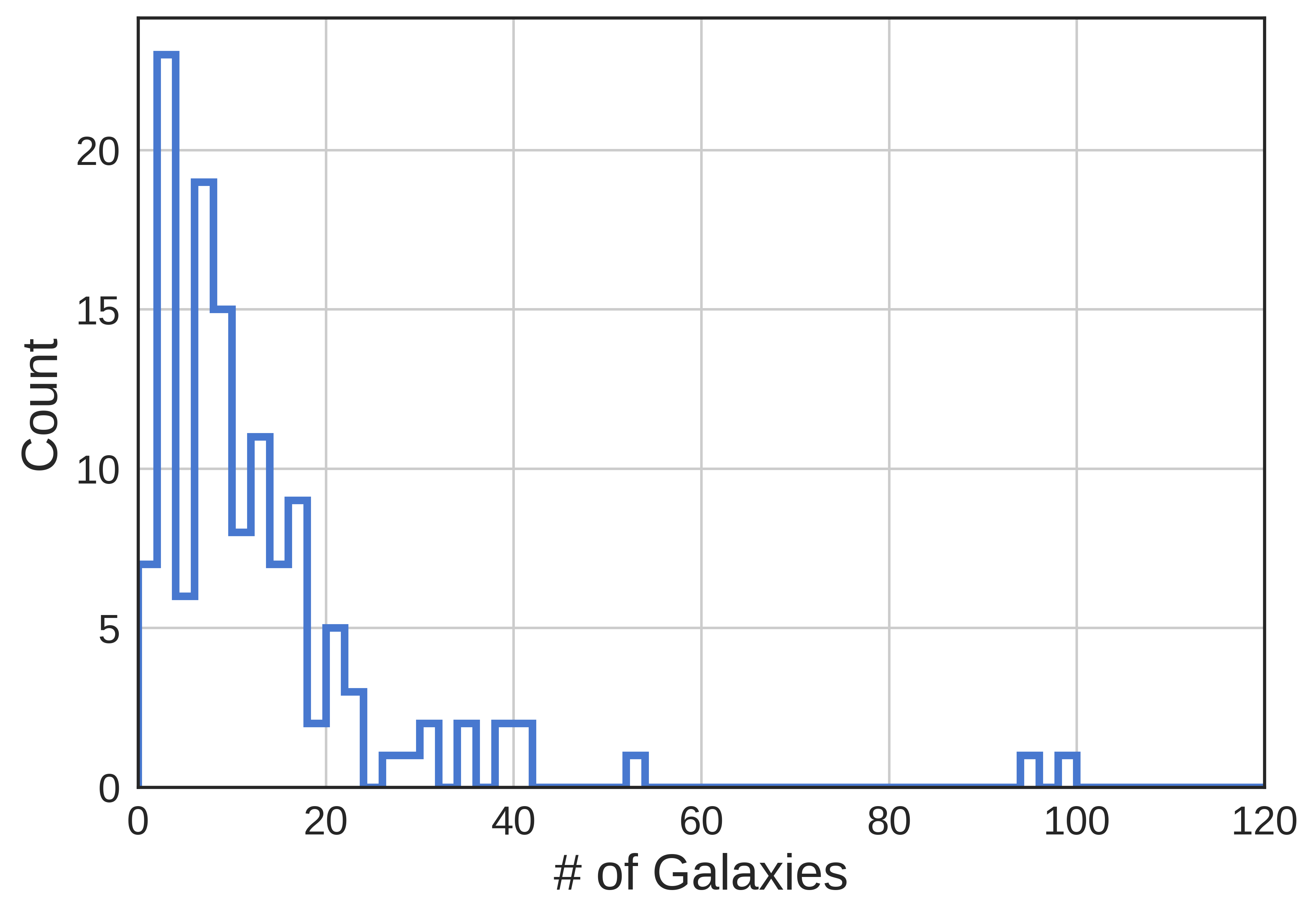}
 \caption{Frequency distribution of the number of spectroscopic members per cluster within $r_{500}$ after removing the high-velocity background component using the velocity cut, equation~(\ref{eq:Vmax}). }
 \label{fig:num-galaxies}
\end{figure}

Figure \ref{fig:num-galaxies} shows the distribution of spectroscopic galaxy counts within $r_{500}$ in the cluster sample after applying the velocity threshold, equation~(\ref{eq:Vmax}).  The modal, median, and mean values are 3, 9, and 12.4 respectively. After applying the velocity and aperture cuts, the main contribution of spectroscopic sample came from GAMA (45\%), VIPERS (30\%), VVDS Deep and Ultra Deep (11\%), SDSS-DR10 (5\%). The remaining catalogs individually contribute less than $2\%$.

In section \ref{sec:sensitivity-analysis}, we investigate the sensitivity of our results to $\vmax$ and $r_{500}$ selection thresholds, not finding statistically significant change.

\section{Cluster Ensemble Velocity Model}\label{sec:like-model}

The study of \citet{Rozo:2015} introduced an ensemble likelihood model for stacked cluster spectroscopy with the goal of assessing the quality of photometric membership likelihoods computed by the redMaPPer cluster finding algorithm \citep{Rykoff:2012}.  This model was designed to take advantage of sparse, wide-area spectroscopic samples, for which each cluster may have only a few member redshifts.  Subsequently, the approach was extended by \citet{Farahi:2016} to infer the scaling of mass with optical richness, $\lambda_{\rm RM}$.  In the present work we follow a similar approach, with X-ray temperature replacing $\lambda_{\rm RM}$. 

\subsection{Ensemble Galaxy Velocity Likelihood}

Power-law scaling relations, originally motivated by the self-similar model \citep{Kaiser:1986}, are confirmed in modern  hydrodynamic simulations, which model baryonic processes in halos  \citep[{\sl e.g.},][]{Truong:2016,McCarthy:2016}.  Consequently, we assume a power-law scaling relation between characteristic galaxy velocity dispersion, $\sigma_{\rm gal}$, and X-ray temperature of the form,
\begin{equation} \label{eq:vel-disp-scaling}
    \sigma_{\rm gal}(\tthreeh,z) \ = \ \sigma_p \ \left( \frac{\ktthreeh}{{\rm k}T_p} \right)^{\alpha} \left( \frac{E(z)}{E(z_p)} \right)^{\beta} , 
\end{equation}
where ${\rm k}T_{p} = 2.2~\kev$ and $z_p=0.25$ are the pivot temperature and redshift, and $E(z) = H(z)/H_0$ is the normalized Hubble parameter.  

The probability distribution function (PDF) of galaxy velocity at a given cluster temperature is taken to be Gaussian with the above dispersion. The ensemble likelihood for the signal component allows for a residual, constant background atop this cluster member signal.  The likelihood for the ensemble cluster-galaxy rest-frame velocity sample is thus
\begin{equation} \label{eq:likelyhoodmodel}
     \mathcal{L} \ = \  \prod\limits_{i=1}^{n} \left[ \ p \ G(v_{{\rm gal,}i} | 0, \sigma_{\rm gal}(T_i,z_i)) + \frac{1-p}{ 2 \vmax(T_i)  } \right]
\end{equation}
where $G$ is the Gaussian distribution with zero mean and standard deviation, $\sigma_{gal}$, $\vgal$ is the line-of-sight (LOS) velocity, equation~(\ref{eq:Velocity}), and the sum $i$ is over all galaxy-cluster pairs in the spectroscopic sample lying below the maximum cutoff, equation~(\ref{eq:Vmax}).  The parameter $p$ is the fraction of galaxies that contribute to the Gaussian component, while $1-p$ is residual fraction of projected systems that are approximated by a uniform distribution in the signal portion of velocity space. 

We maximize this likelihood with respect to the four model parameters, $\sigma_p$, $\alpha$, $\beta$, and $p$.  Below we find that the redshift evolution parameter, $\beta$, is both relatively poorly constrained and consistent with zero.  We therefore also perform a restricted analysis in which we assume self-similar evolution (SSE), with $\beta = 0$.  

\subsection{Ensemble Velocity Model in Simulations}

This model has been tested against simulation by \citet{Farahi:2016}, using cluster richness instead of X-ray temperature, with several key findings. First, the spectroscopic mass estimate is a nearly unbiased estimator of $\langle \ln M_{\rm mem} | \lambda_{\rm RM} \rangle$, where $M_{\rm mem}$ is the mass of the underlying halo that contributes the maximum fraction of the cluster's photometric member galaxies assigned by redMaPPer.  Second, galaxies lying in the signal region consist of a majority coming from the top-ranked, member-matched halo ($\sim 60\%$) as well as locally projected galaxies ($\sim 40\%$) lying outside the matched halo.  
Finally, the main source of systematic uncertainty in the SDSS cluster mass estimate of \citet{Farahi:2016} is uncertainty in the magnitude of the galaxy velocity bias.

\section{Velocity Scaling Results}\label{sec:results}

\begin{table}
\centering
\setlength\extrarowheight{2.5pt}
\caption{Expectation values and standard deviations of the marginalized posterior distributions of free parameters of the model defined in equations~(\ref{eq:vel-disp-scaling}) and (\ref{eq:likelyhoodmodel}).  Parameters listed below are for the fiducial model; the self-similar evolution model, with $\beta$ set to zero, returns identical central values and errors for the other parameters and so are not listed.  } 
\label{tab:fiducialModel}
\begin{tabular}{|c|c|c|c|}
\hline
$\sigma_p ~ [{\rm km/s}]$ & $\alpha$ & $\beta$ & $p$ \\ 
\hline
 $539 \pm 16$ & $0.63 \pm 0.05$ & $-0.49 \pm 0.38$ & $0.88 \pm 0.015$ \\ 
\hline
\end{tabular}
\end{table}

In this section, we present the inferred $\sigma_{gal} - \ktthreeh$ scaling relation for the full cluster sample.  The fiducial analysis uses the signal velocity threshold of equation~(\ref{eq:Vmax}), an angular limit of $r_{500}$, and 
solves for the four degrees of model freedom using the entire sample.  
Sensitivity tests of the angular and velocity thresholds used in our fiducial treatment are presented in the next section. 

We run the MCMC analysis module \texttt{PyMC} \citep{PyMC:2010} to maximize the likelihood and recover the scaling relation parameters between velocity dispersion of galaxy members and temperature of hot cluster gas.
We assume a uniform priors on all parameters, with the following domain limits: $p \in [0, 1]$, $\sigma_p \in [50, 1000] \kms$, $\alpha \in [-10, 10]$, and $\beta \in [-10, 10]$.

The best-fit parameter values for the fiducial model and the restricted SSE model are given in Table~\ref{tab:fiducialModel}.  
The posterior PDFs of the free parameters are presented in Appendix \ref{app:posterior}.  

For the fiducial treatment, the posterior constraint on the slope of galaxy velocity dispersion scaling with temperature is $\alpha = 0.63 \pm 0.05$,  is  in tension with the self-similar expectation of $0.5$.  A slope steeper than self-similar could potentially arise from AGN feedback effects on the ICM. Recent simulations including AGN feedback exhibit shifts in the global ICM temperature of halos that are mass-dependent, with larger increases seen at lower masses \citep{Lebrun:2016,Truong:2016}.  Since the galaxy velocity dispersion is not directly coupled to AGN activity, the impact on the ICM would lead to $\alpha > 0.5$.

We find no significant change in the scaling amplitude with redshift but our constraint is weak, $\beta = -0.49 \pm 0.38$.  Since the fiducial analysis yields no evidence of redshift evolution, it is no surprise that the posterior SSE parameter values are identical to those of the fiducial analysis.  

The Gaussian component amplitude, $p$, is close to, but significantly different from unity.   While the value of $0.88 \pm 0.02$ is consistent with the $0.916 \pm 0.004$ value found by \citet{Rozo:2015} in their study of SDSS redMaPPer clusters, differences in selection and measurement preclude a direct comparison.  Besides sample selection differences, the SDSS galaxy velocities are pairwise with respect to the central galaxy's velocity, whereas ours are determined by the mean cluster redshift, $z_c$.  Some of the difference could reflect mis-centering, as a larger fraction of mis-centered clusters both reduces $p$ and increases $\sigma_p$ \citep{Farahi:2016}. We defer detailed modeling of such selection effects to future work. 

\begin{figure}
\centering
\includegraphics[width=1.0\linewidth]{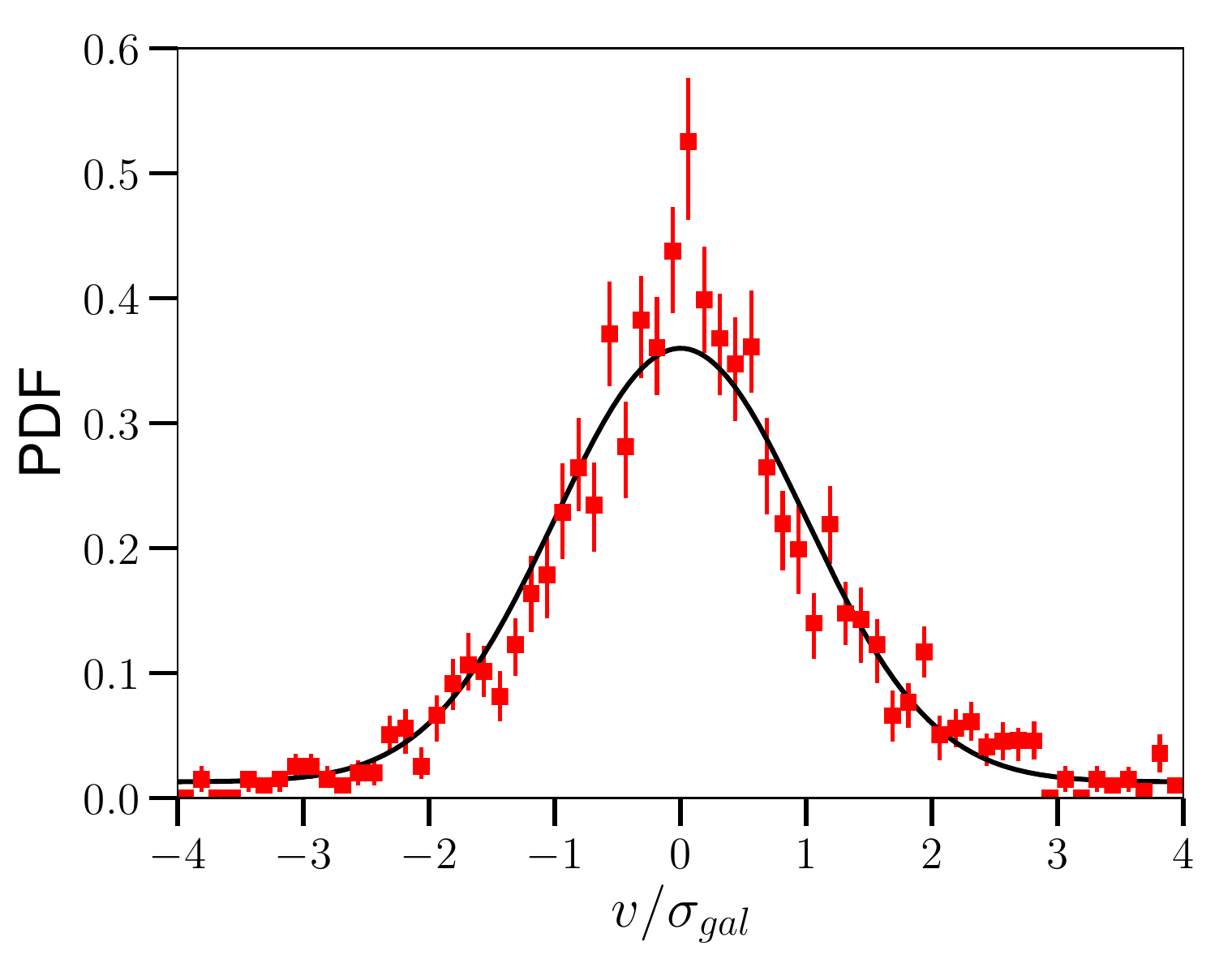}
\caption{Normalized residuals of galaxy velocity about the mean scaling relation in the fiducial analysis.  Red points show the data and the black line is the model, equation~(\ref{eq:likelyhoodmodel}), a mixture of a Gaussian and a uniform distribution.  Error bars are calculated by bootstrapping the velocities of the spectroscopic sample, using $64$ bins between $-4$ and $4$ in $\vgal /\sigma_{gal}$.  See text for discussion of the goodness of fit.}
\label{fig:goodness-of-fit}
\end{figure}

Normalized velocity residuals about the mean scaling behavior in the fiducial analysis are shown in Figure~\ref{fig:goodness-of-fit}.  
We bootstrap the galaxy sample to compute means and standard deviations of the PDF in 64 bins between $-4$ and $4$ in 
$v/\sigma_{gal}$, and these are shown as points with error bars in the figure. The black line is the model, a Gaussian of zero mean, unit variance and amplitude given by the fiducial best fit plus a constant background.

From Figure~\ref{fig:goodness-of-fit}, it is evident that our fit is \emph{not} a good fit to data in the standard chi-squared sense.
The normalized velocity PDF structure is very similar to that seen by \citet{Rozo:2015} and \citet{Farahi:2016} for redMaPPer clusters and simulations, respectively.
We find $\chi^2 / {\rm dof} = 74 / 44$ for $\vgal / \sigma_{gal} \in [-3,3]$, which is less than that for the best-fit value found by \citet{Rozo:2015} for SDSS redMaPPer clusters, $\chi_{\rm SDSS}^2 / {\rm dof} = 96 / 26$. 

While the centrally peaked nature of the normalized velocity PDF remains to be carefully modeled, two potential sources are likely to be important.  One is projected large-scale structure; the \citet{Farahi:2016} simulations show that only $\sim 60\%$ of the galaxies in the signal component of velocity space actually lie within $r_{200}$ of the halo matched to each member of the cluster ensemble.  Another is intrinsic scatter in $\sigmagal - T_X$, which will distort the Gaussian shape. The fact that the $\chi^2 / {\rm dof}$ is smaller for the XXL sample compared to SDSS redMaPPer may reflect the fact that the intrinsic scatter in galaxy velocity dispersion is smaller at fixed temperature than at fixed richness, but differences in selection may also play a role.

Although the best fit is not a good fit to a Gaussian, the simulation of \citet{Farahi:2016} show that the derived galaxy velocity dispersion scaling is unbiased with respect to the log-mean value obtained by matching each cluster to the halo that contributes the majority of its galaxy members. Because the galaxy velocities in that simulation are unbiased relative to the dark matter by construction, the virial mass scaling derived from the galaxy velocity dispersion, $M(\lambda_{\rm RM},z) \propto \sigma_p^3 (\lambda_{\rm RM},z)$, presents an unbiased estimate of the log-mean, membership-matched halo mass of the cluster ensemble.  The reader interested primarily in mass scaling estimates can move directly to \S\ref{sec:massCalibration}.  

We turn next to comparing our scaling of galaxy velocity dispersion with gas temperature to previous work, and then explore the robustness of our parameter values in \S\ref{sec:sensitivity-analysis}.

\subsection{Comparison with Previous Studies}

\begin{table*}
\centering
\setlength\extrarowheight{2.5pt}
\caption{Summary of published $\sigma_{gal} -  \kt$ scaling relation parameters, using the notation$^1$ of equation~(\ref{eq:vel-disp-scaling}) } \label{tab:comparisonObsScaling} 
\begin{threeparttable}
\begin{tabular}{|c|c|c|c|c|c|c|}
\hline
Source  &  $\sigma_p ~ ({\rm km\ s^{-1}})$  &  $\alpha$  &  $\beta$  &  fitting method  & $N$ & redshift \\ \hline
\hline
This work & $539 \pm 16$ & $0.63 \pm 0.05$ & $-0.49 \pm 0.38$ & Ensemble ML & 132 & $ z < 0.6$ \\ \hline
\citet{Wilson:2016} & $497 \pm 85$ & $0.86 \pm 0.14$ & $-0.37 \pm 0.33$ & ODR~\tnote{2} & 38 & $z < 1$ \\ \hline
\citet{Nastasi:2014} & $508 \pm 147$ & $0.64 \pm 0.34$ & - & BCES bisector & 15 & $ 0.64 \le z \le 1.46$ \\ \hline
\citet{Xue:2000} & $523 \pm 13$ & $0.61 \pm 0.01$ & - & ODR~\tnote{2} & 145 & $z < 0.2$ \\ \hline
\end{tabular}
\begin{tablenotes}\footnotesize
\item[1] Note that sample definitions, analysis methods and notation vary across sources. Published intercepts are renormalized to the fixed pivot temperature and redshift used in equation~(\ref{eq:vel-disp-scaling}).
\item[2] Orthogonal Distance Regression
\end{tablenotes}
\end{threeparttable}
\end{table*}

Soon after early observations of extended X-ray emission from clusters indicated a thermal gas atmosphere, a dimensionless parameter of interest emerged: the ratio of specific energies in galaxies and hot gas, 
$\betaspec =  \sigmagal^2 / (\kt/\mu m_p) $, where $\mu$ is the mean molecular weight of the plasma and $m_p$ is the proton mass (note this beta is fundamentally different from the symbol used in \S \ref{sec:like-model}). 

Early estimates of this ratio in small observational samples \citep{Mushotzky:1978} and gas dynamic simulations \citep{Evrard:1990, Navarro:1995} yielded $\betaspec \approx 1$, consistent with a scenario in which both components are in virial equilibrium within a common gravitational potential.  More recently, this ratio has been explored at high redshift; 
\citet{Nastasi:2014} find $\betaspec = 0.85 \pm 0.28 $ for 15 clusters with $z>0.6$.

\begin{figure}
\centering
\includegraphics[width=0.95\linewidth]{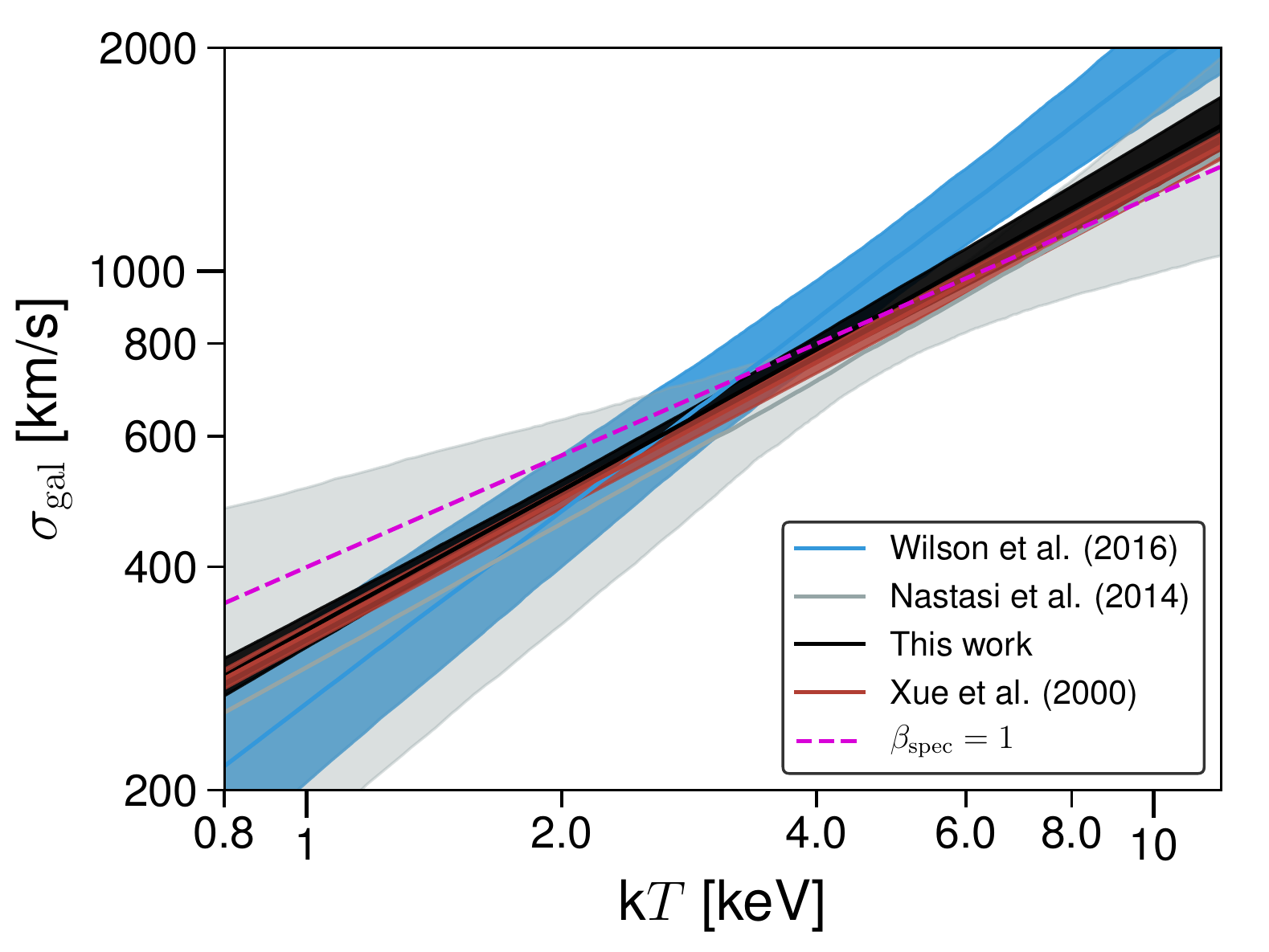}
\caption{Comparison of the $\sigmagal - \ktthreeh$ scaling relation of this work with prior literature, as labeled. Shaded regions are $1 \sigma$ uncertainty on the expected velocity dispersion at given temperature. The magenta line is the locus of constant specific energy ratio, $\betaspec = \sigmagal^2/(\kt / \mu m_p)  = 1$ with $\mu=0.6$. The slope of \citet{Wilson:2016} suffers from a potential bias discussed in the text.}
\label{fig:sig-T-scaling-relation}
\end{figure}

Figure \ref{fig:sig-T-scaling-relation} compares the fiducial scaling relation of this work to previous determinations in the literature.  In addition, the dashed (magenta) line shows $\betaspec =1$ assuming mean molecular weight $\mu = 0.6$, appropriate for a metal abundance of $0.3 Z_{\sun}$. 
Shaded regions show $1 \sigma$ uncertainty on the expected velocity dispersion at a given temperature.

Table~\ref{tab:comparisonObsScaling} summarizes the comparison with previous studies. The published scaling relations are re-evaluated at the pivot point of this work to be directly comparable. When appropriate, errors in the published slope are propagated to the normalization error.  

 The measured slope is consistent between our work and previous works.  \citet{Wilson:2016} find a slope $0.86 \pm 0.14$ for a sample of 38 clusters from the XMM Cluster Survey.  Using simulations, however, they show that the orthogonal fitting method on their sample produces a  substantial overestimate in slope, by $\sim 0.3$, in the test shown in their Table 7 and Figure 9.  They caution that their fit overestimates the velocity dispersion of clusters above $5~\kev$. Similarly \citet{Ortiz-Gil:2004} uses the orthogonal fitting method and find a steep slope $\sim 1.00 \pm 0.16$ for a sample of 54 clusters. 

If a bias correction is applied, the slope of \citet{Wilson:2016} reduces to $\sim 0.55$, consistent with our findings.  
We note that a smaller shift of $\sim 0.2$ would bring the \citet{Ortiz-Gil:2004} result into consistency with self-similarity at the $2\sigma$ level.  For a heterogeneous sample constructed from the literature, \citet{Xue:2000} report a slope of $0.61 \pm 0.01$, consistent with our result.

The velocity dispersion normalizations given in Table~\ref{tab:comparisonObsScaling} at the pivot temperature and redshift are all in good agreement within their stated errors. The $3\%$ fractional uncertainty in our quoted normalization is among the tightest published constraints, comparable to the statistical error of the more heterogeneous sample of \citet{Xue:2000}. 

\section{Systematic Errors and Sensitivity Analysis} \label{sec:sensitivity-analysis}

In this section, we investigate sources of uncertainty in the scaling presented in the previous section, including survey selection and the sensitivity of the posterior parameters to the details of the spectroscopic sample used to define the signal region.

\begin{table*}
\setlength\extrarowheight{2.5pt}
\centering
\caption{Sensitivity analysis of $\sigmagal - \ktthreeh$ inferred parameters. See text for further discussion.\label{tab:sensitivityAnalysis}} 
\begin{threeparttable}
\begin{tabular}{|c|c|c|c|c|c|c|}
\hline
Model  &  $\sigma_p ~ [{\rm km/s}]$  &  $\alpha$  &  $\beta$  &  $p$  & \# Clusters & \# Galaxies             \\ \hline
\hline
Fiducial & $539 \pm 16$ & $0.63 \pm 0.05$ & $-0.43 \pm 0.38$ & $0.88 \pm 0.02$ & 128  & 1592 \\ \hline
Measured $\ktthreeh$ only & $547 \pm 17$ & $0.60 \pm 0.05$ & $-0.39 \pm 0.39$ & $0.87 \pm 0.02$ & 103 & 1421  \\ \hline
$ r < 0.5 r_{500} $ & $509 \pm 20$ & $0.67 \pm 0.07$ & $-1.29 \pm 0.50$ & $0.90 \pm 0.02$ & 127 & 891 \\ \hline
$ r < 2.0 r_{500} $ & $557 \pm 13$ & $0.56 \pm 0.04$ & $0.42 \pm 0.32$ & $0.82 \pm 0.02$ & 131 & 2810 \\ \hline
$v_{max} = 2000 \kms$ ~\tnote{1} & $526 \pm 18$ & $0.62 \pm 0.05$ & $-0.50 \pm 0.40$ & $0.88 \pm 0.02$ & 128 & 1557 \\ \hline
$v_{max} = 3000 \kms$ ~\tnote{1} & $549 \pm 15$ & $0.63 \pm 0.05$ & $-0.45 \pm 0.37$ & $0.88 \pm 0.02$ & 128 & 1617 \\ \hline
$\alpha_{V_{max}}  = 0.3$~\tnote{2} & $539 \pm 16$ & $0.61 \pm 0.05$ & $-0.46 \pm 0.39$ & $0.88 \pm 0.02$ & 128 & 1591 \\ \hline
$\alpha_{V_{max}} = 0.7$~\tnote{2} & $543 \pm 16$ & $0.65 \pm 0.05$ & $-0.48 \pm 0.38$ & $0.88 \pm 0.02$ & 128 & 1589 \\ \hline
$z_c ~\tnote{3}~ > 0.25$ & $550 \pm 32$ & $0.58 \pm 0.09$ & $-0.82 \pm 0.79$ & $0.87 \pm 0.02$ & 84 & 814 \\ \hline
$z_c ~\tnote{3}~ \le 0.25$ & $576 \pm 48$ & $0.63 \pm 0.06$ & $0.63 \pm 1.42$ & $0.88 \pm 0.02$ & 44 & 778 \\ \hline
\end{tabular}
\begin{tablenotes}\footnotesize
\item[1] Normalization of the maximum velocity threshold in equation~(\ref{eq:Vmax})
\item[2] Slope in temperature of the maximum velocity threshold in equation~(\ref{eq:Vmax})
\item[3] Cluster redshift. 
\end{tablenotes}
\end{threeparttable}
\end{table*}

Table~\ref{tab:sensitivityAnalysis} summarizes the results of the tests presented below.  A cursory look at the table indicates that most parameters shift by modest amounts, typically within one or two standard deviations of the fiducial result, with the exception of the Gaussian amplitude, $p$, discussed further below.   

\subsection{Temperature Estimates}

As presented in \S\ref{sec:temperatures}, the XXL temperatures are directly determined for 103 of the 128 clusters in our sample.  A natural question to ask is whether our results are sensitive to the temperature estimation method applied to the remaining 25 clusters.  

We first note that the 103 systems with measured $\tthreeh$ tend to be more massive at a given redshift, with higher galaxy richness.  The higher richness translates into more galaxies with spectroscopy, and it turns out that this subset holds most of the statistical weight of the spectroscopic sample.  Within the fiducial $r_{500}$ aperture, there are 1421 galaxies in the 103 clusters with direct temperatures, compared with 171 galaxies in the 25 clusters with inferred temperatures.  So $\sim 90\%$ of the statistical weight comes from clusters with measured temperatures.  

As a consistency check, we refit the scaling relation after removing all clusters with inferred temperature from the sample.  The parameter constraints remain consistent with our fiducial analysis.

\subsection{Angular Aperture}

The velocity dispersion of dark matter particles in simulations varies weakly as a function of distance from the halo center \citep[][]{Old:2013}, and this effect has been confirmed observationally \citep{Biviano:2003}.
We test the sensitivity of our fit parameters by varying the angular aperture of inclusion by factors of $2^{\pm 1}$ from the fiducial value of $r_{500}$.  Note that the size of the sample varies slightly as the aperture is changed.  The main change is that a larger aperture induces a larger projection effect, evident from the Gaussian normalization, $p = 0.82 \pm 0.02$ for $2r_{500}$ versus $p = 0.90 \pm 0.02$ for $0.5 r_{500}$. There are modest trends in the other parameters, including a slightly steeper slope $\alpha = 0.67 \pm 0.07$ at $0.5 r_{500}$, and $\beta$ is not consistent with $0$ at the $\sim 2\sigma$ level at $0.5r_{500}$, but the statistical power of the sample is insufficient to determine these trends with high precision.

\subsection{Signal Component Maximum Velocity}

 Recall that the likelihood model is applied to a subset of all spectroscopic galaxies that lie in the signal region, with rest-frame velocities below a maximum value, $\vmax(\tthreeh)$, given by equation~(\ref{eq:Vmax}). We test the effect of this maximum by independently varying the amplitude by $\pm 500 \kms$ (or $\pm 20\%$) and the power-law index by $\pm 0.2$. The number of signal galaxies does not vary much with these changes, indicating that our fiducial cut is roughly identifying the caustic edge that separates bound and unbound galaxies in clusters \citep{Miller:2016}.  All parameters remain within $1\sigma$ of their fiducial values as these changes are made.  

\subsection{Redshift Range}

We take the pivot redshift in this work, $z_p=0.25$, and split the full sample into high and low redshift subsets.
For these, we do not find statistically significant deviations from the fiducial model parameters.  The changes in the normalization, slope, redshift evolution, and parameter $p$ are all less than $1 \sigma$.
Although, as to be expected, there remains no effective constraints on the redshift evolution factor.

\subsection{X-ray Selection and Malmquist Bias} \label{sec:selection-bias}

The aim of our analysis is to produce unbiased estimates of the scaling relations inherent to the population of dark matter halos.  Selection by X-ray flux and angular size \citep{Pacaud:2006} can introduce bias in the inferred $\sigmagal-\ktthreeh$ scaling relation if there is non-zero covariance between X-ray selection properties and galaxy velocity dispersion \citep[see \S5.1 in ][]{Kelly:2007}. Such data sets are said to be ``truncated'', and the truncation effects need to be explicitly modeled in the likelihood.

There have not yet been observational estimates of the correlation between galaxy velocity dispersion and X-ray properties at fixed halo mass.  Halos in the Millennium Gas simulations of \citet{Stanek:2010} show intrinsic correlation coefficients of $\sim 0.3$ for $L_X$ and $\sigma_{\rm DM}$, where $\sigma_{\rm DM}$ is the velocity dispersion of dark matter particles in the halos. 
However, translating this estimate into correlations involving $\sigmagal$ projected along the line-of-sight is non-trivial and lies beyond the scope of this work. Redshift-space projection presumably dilutes any intrinsic halo correlation, unless the source of the projected velocity component also carries associated X-ray emission. 

The magnitude of potential selection biases can be addressed by simulating the entire process of survey selection and subsequent spectroscopic analysis, along the lines of that done by \citet{Farahi:2016} for redMaPPer optical selection.  We defer that work to future analysis. From the perspective of halo mass estimation, corrections to the velocity dispersion scaling from sample selection are likely to be smaller than the systematic uncertainty associated with galaxy velocity bias, as discussed below.


\section{Ensemble Dynamical Mass Scaling of XXL Clusters} \label{sec:massCalibration}

As previously noted, \citet{Farahi:2016} use sky realizations derived from lightcone outputs of cosmological simulations to show that the mass determined through virial scaling of the ensemble, or stacked, pairwise velocity dispersion offers an unbiased estimate of the log-mean mass of halos matched via joint galaxy membership.  Here, we apply this approach to the fiducial velocity dispersion scaling in order to  estimate the characteristic mass scale, $\langle \ln M_{200}|T_X \rangle$  of XXL clusters as a function of temperature at the pivot redshift, $z_p = 0.25$.

The simulation of \citet{Farahi:2016} assumed galaxies to be accurate tracers of the dark matter velocity field, but real galaxies may be biased tracers. To estimate the velocity dispersion of the underlying dark matter from the galaxy redshift measurements, we introduce a velocity bias factor, $b_v$, defined as the mean ratio of galaxy to dark matter velocity dispersion within the target projected $r_{200}$ region used in our analysis. The normalization of the dark matter velocity scaling with temperature is then 
\begin{equation}
	\sigma_{p,DM} = \frac{\sigma_p}{b_v}, 
\end{equation}
where $\sigma_p$ is the galaxy normalization with temperature, equation~(\ref{eq:vel-disp-scaling}).  

Following \citet{Farahi:2016}, we proceed by: i) imposing an external $b_v$ estimate to derive the normalization of the {\sl dark matter} virial velocity scaling with X-ray temperature, then ii) applying the dark matter virial relation calibrated by \citet{Evrard:2008} to determine the scaling of total system mass with temperature.  

We use $b_v = 1.05 \pm 0.08$ which is an empirical estimate derived from redshift-space clustering of bright galaxies by \citet{GuoII:2015}.  A similar value of $1.06 \pm 0.03$ is found in the simulation study of \citet{Wu:2013}, although that study found galaxy bias slightly below 1 for the brightest galaxies.   We note that the peak of distribution of absolute r-band magnitude of selected galaxies in this work is ${\rm M}_{r} = 21.5$ (see appendix \ref{app:r-band}), which is consistent with the brightest galaxy sample of \citep{GuoII:2015}.  Using a velocity bias of $1.05 \pm 0.08$ leads to an estimate of the dark matter velocity dispersion at the pivot temperature and redshift, 
\begin{equation}
 	\sigma_{p,DM} = 516 \pm 43~ {\rm km/s} .
\end{equation}
Note that $\sigma_{p,DM}$ uncertainty has contribution from the $b_v$ prior and $\sigma_p$ posterior.  

The virial scaling of halos in simulations displays a linear relationship between the cube of the dark matter velocity dispersion, $\sigma^3_{p,DM}$, and a mass measure, $E(z) M_\Delta$, where $E(z) = H(z)/H_0$ is the normalized Hubble parameter.  
Using equation~(6) and Table 3 of \citet{Evrard:2008} along with $h=0.7$, the total mass within $r_{200}$ at the pivot temperature and redshift is 
\begin{equation}
	\langle \ln (M_{200} /10^{14}~\msun )\rangle = 0.33 \pm 0.24 , 
\end{equation}
corresponding to $M_{200} = (1.39^{+0.37}_{-0.30}) \times 10^{14}~\msun$. 

The full velocity scaling implies a log-mean mass for the XXL selected cluster sample of 
\begin{equation}
    \left\langle \ln \left(\frac{E(z) M_{200}}{10^{14}\msun}\right) | T , z \right\rangle = \pi_{T} + \alpha_{T} \ln \left(\frac{T}{T_p} \right) + \beta_{T} \ln \left(\frac{E(z)}{E(z_p)} \right)
\end{equation}
with intercept $\pi_{T} = 0.45 \pm 0.24$, temperature slope $\alpha_{T} = 3\alpha = 1.89 \pm 0.15$, redshift slope $\beta_{T} = 3\beta = -1.29 \pm 1.14$.  Recall that this result is based on 300 kpc temperature estimates, $T \equiv \tthreeh$. 

\citet{Biviano:2006} have examined the robustness of virial mass estimates in a cosmological hydrodynamic simulation. 
They find that dynamical mass estimates are reliable for densely sampled clusters (over 60 cluster members). 
Due to the ensemble technique adapted here, this work does not suffer from sparse sampling of cluster members.
Generally speaking, stacking techniques reduce the noise associated with sparse samples, at the price of not constraining the intrinsic scatter. 

While we explicitly remove extreme projected outliers in velocity space (see Fig.~\ref{fig:vel_structure}) and account for a residual, constant contribution in the velocity likelihood, it is worth noting that the central Gaussian component has contributions from galaxies that do not lie in the main source halo. While this component retains some degree of projected galaxies, \citet{Farahi:2016} shows that the dynamically-derived mass is a robust estimate of log-mean mass at a given observable, in that case $\langle \ln M_{200} | \lambda_{\rm RM}, z \rangle$.  While the optical and X-ray samples are selected differently, not enough is known about hot gas and galaxy property covariance to model selection effects precisely.  We discussed in \S\ref{sec:selection-bias} why  selection effects are unlikely to imprint significant bias into the inferred scaling relation.

\subsection{Comparison with Previous Studies}

\begin{figure}
\centering
\includegraphics[width=1.0\linewidth]{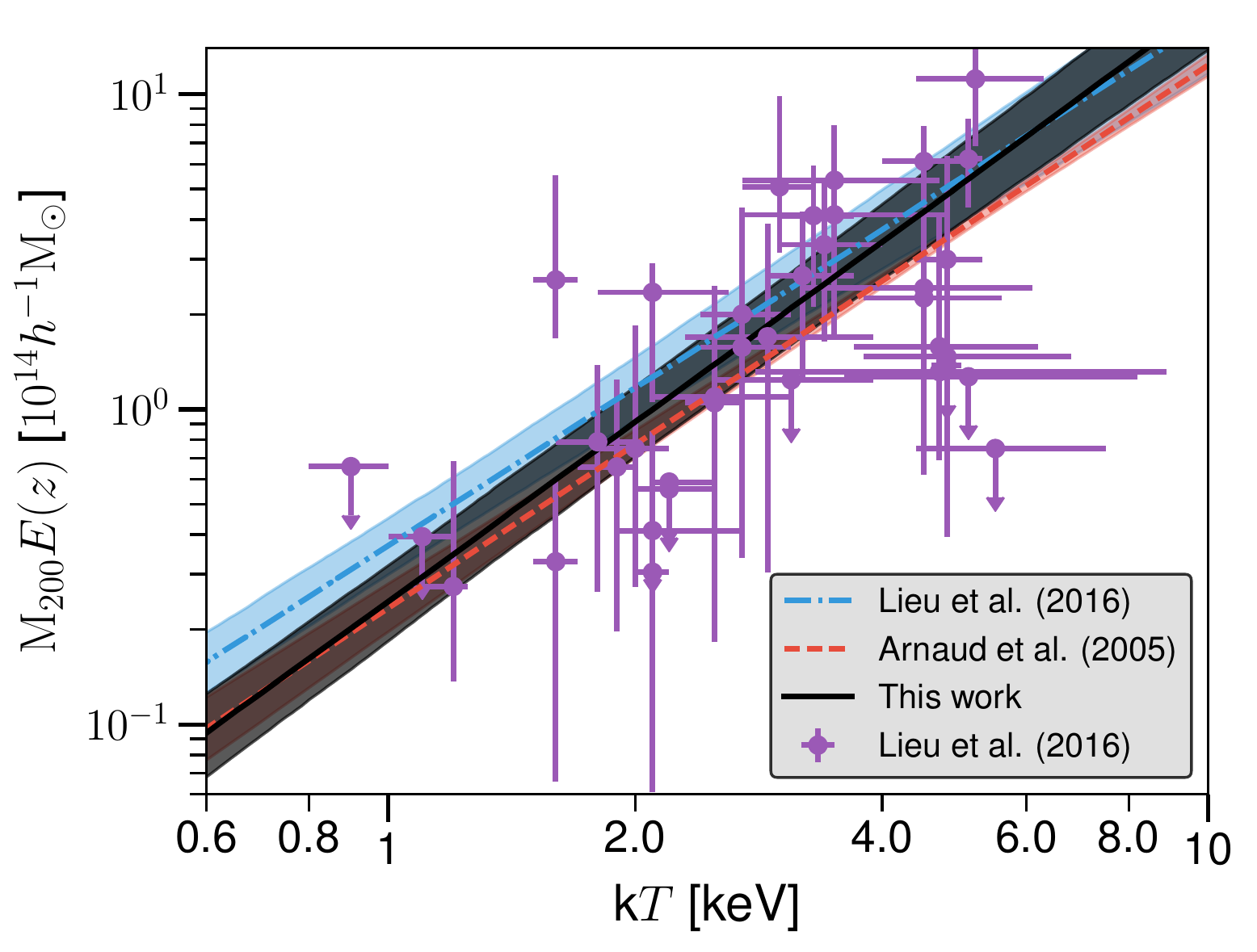}
\caption{The $M_{200} - {\rm k}T$ scaling relation from this work (black line and dark shaded region) is compared with published relations given in the legend and Table~\ref{tab:comparison}. Shaded regions are the $1 \sigma$ uncertainty in the expected mass at a given temperature. See the text for more discussion.}
\label{fig:mass-T-scaling-relation}
\end{figure}

Figure~\ref{fig:mass-T-scaling-relation} compares the mass-temperature scaling relation, a dynamical mass estimates, derived in this work with previous studies that use weak lensing \citepalias{Lieu:2016} and hydrostatic \citep{Arnaud:2005} mass estimates.  
Overall, there is a good agreement within the uncertainties.  

The data points with error bars are weak lensing estimates of $M_{200}$ for a subsample of the 100 brightest clusters in XXL \citepalias{Lieu:2016}. 
In order to directly compare our MOR with \citetalias{Lieu:2016} and other works, we evaluate all results at $z=0$ using $h=1$.  When shifting the normalization, we assume SSE, $\beta_T=0$, yielding $\pi_T = 0.09 \pm 0.25$.
  
Assuming self-similar redshift evolution, \citetalias{Lieu:2016} estimated the mass - temperature scaling relation using a subsample of 38 out of 100 brightest XXL clusters. To improve their constraint, their sample is complemented with weak lensing mass measurements from clusters in the COSMOS \citep{Kettula:2013} and CCCP \citep{Hoekstra:2015} cluster samples.  While the data points plotted in Figure~\ref{fig:mass-T-scaling-relation} are taken directly from \citetalias{Lieu:2016}, their published MOR is framed in terms of $M_{500}$.  We therefore convert the normalization to  $M_{200}$ using 
an NFW profile with concentration $c = 3.1$, the median value of the \citetalias{Lieu:2016} sample, for which $M_{200}/M_{500} = 1.4$.  The slope of the weak lensing relation lies within $\sim 1 \sigma$ of the self-similar expectation of $1.5$.   

The assumption of hydrostatic equilibrium is commonly used to derive masses from X-ray spectral images, and 
\citet{Arnaud:2005} apply this method to a sample of ten nearby, $z < 0.15$, relaxed clusters in the X-ray temperature range $[2-9]~{\rm keV}$.
The masses are derived from NFW fits to the mass profiles, obtained under the hydrostatic assumption using measurements from the XMM-Newton satellite.  Note that they use a core-excised spectroscopic temperature from a $0.1 r_{200} \le r \le 0.5 r_{200}$ region.
Our result is consistent with that of \citet{Arnaud:2005} within their respective errors. 

\citet{Kettula:2015} combine 12 low mass clusters from the CFHTLenS and XMM-CFHTLS surveys with 48 high-mass clusters from CCCP \citep{Hoekstra:2015} and 10 low-mass clusters from COSMOS \citep{Kettula:2013}.  From this sample of 70 systems, 
they measure 
a mass - temperature scaling relation with slope $1.73 \pm 0.19$ for $M_{200}$.  When $M_{500}$ is used, they find a slope of $1.68 \pm 0.17$ which they argue may be biased by selection.  Applying corrections to this (Eddington) bias, they find a slope of $1.52 \pm 0.17$, consistent with self-similarity.  

\begin{table*}
\setlength\extrarowheight{2.5pt}
\centering
\caption{Comparison of the mass normalization, 
$\ln A = \langle \ln (M_{200}/10^{14}\hinv \msun) \, | \, \kt = 2.2 \kev, z=0 \rangle$, and slope of the mass--temperature determined by the works listed. \label{tab:comparison}} 
\begin{threeparttable}
\begin{tabular}{|c|c|c|c|c|c|c|}
\hline
Paper  &  $\ln A$  & Slope  &  Mass Proxy  & Number of Clusters & redshift \\ \hline
\hline
This work & $0.09 \pm 0.25$ & $1.89 \pm 0.15$ & Dynamical Mass & 132 & $z < 0.6$ \\ \hline
\citetalias{Lieu:2016}~\tnote{1} & $0.31 \pm 0.23$ & $1.67 \pm 0.14$ & Weak-lensing Mass & 96 & $0.1 < z < 0.6$ \\ \hline
\citet{Kettula:2015}~\tnote{2} & $0.43 \pm 0.17$ & $1.73 \pm 0.19$ & Weak-lensing Mass & 70 & $0.1 \le z \le 0.5$ \\ \hline
\citet{Arnaud:2005}~\tnote{3} & $-0.09 \pm 0.09$ & $1.72 \pm 0.10$ & Hydrostatic Mass & 10 & $z < 0.16$ \\ \hline
\end{tabular}
\begin{tablenotes}\footnotesize
\item[1] The normalization is converted from $M_{500}$ to $M_{200}$ as described in the text.
\item[2] CFHTLenS + CCCP + COSMOS cluster sample.
\item[3] Spectroscopic temperature within the $0.1 r_{200} \le r \le 0.5 r_{200}$ region. All clusters.
\end{tablenotes}
\end{threeparttable}
\end{table*}

Table \ref{tab:comparison} summarizes these comparisons, showing the slopes and normalizations scaled to $z=0$ for a pivot X-ray temperature of $2.2 \kev$.
The expected log mass is the largest for weak-lensing proxies, and smallest under the  hydrodynamic assumption, but they are statistically consistent within their stated $10-20\%$ errors.
The slope derived in this work is statistically consistent with the scalings derived from weak-lensing and hydrostatic techniques.  In agreement with prior work, we find a significantly ($>2.5 \sigma$) steeper slope than the expected self-similar value of $1.5$.
A more precise comparison would need to take into account different approaches to measuring X-ray temperature, as well as potential instrument biases \citep{Zhao:2015,Schellenberger:2015}. 
For example, \citet{Arnaud:2005} and \citet{Kettula:2015} measure core-excised temperatures within $r_{200}$ while the temperatures used in this work are measured within fixed physical radius.
Comparing the non-core excised temperatures of XXL clusters with the core excised temperatures used by \citet{Kettula:2013}, \citetalias{Lieu:2016} found a mean ratio of $\langle T_{300~\kpc} / T_{0.1-0.5r_{500,WL}} \rangle = 0.91 \pm 0.05$. 

 Several independent hydrodynamic simulations that incorporate AGN feedback, including models from variants of the Gadget code \citep[cosmo-OWLS; ][]{Lebrun:2016,Truong:2016} as well as RAMSES Rhapsody-G \citep{Hahn:2015}, find slopes near $1.7$ for the scaling of mean mass with spectroscopic temperature.  These results are in agreement with our finding. We note that the cluster sample used in this work is dominated by systems with $\kt < 3~{\rm keV}$, while \citet{Lieu:2016}'s cluster sample is dominated by clusters with temperature above $3~{\rm keV}$. Slopes steeper than the self similar prediction for low temperature systems have been noted in preceding observational works as well \citep[{\sl e.g.},][]{Arnaud:2005,Sun:2009,Eckmiller:2011}. 

\subsection{Velocity bias}

As in the original application of \citet{Farahi:2016} to estimate the mass-richness scaling of redMaPPer clusters, the dominant source of systematic uncertainty in ensemble dynamical mass estimates comes from the uncertainty in the velocity bias correction.

Dynamical friction is a potential physical cause for the velocity bias that would generally drive galaxy velocities to be lower than that of dark matter particles within a halo \citep[{\sl e.g.}][]{Richstone:1975, Cen:2000,Yoshikawa:2003}.
On the other hand, clusters that are undergoing mergers tend to have galaxy members with a larger velocity dispersion relative to the dark matter particles \citep{Faltenbacher:2006}, and merging of the slowest galaxies onto the central galaxy could also tend to drive $b_v$ to be greater than one.  
 These  competing effects are subject to observational selection in magnitude, color, galaxy type, star formation activity and aperture which need to be addressed with larger sample size.  
 There is growing observational evidence that velocity bias is a function of the aforementioned selection variables \citep[{\sl e.g.}, ][]{GuoII:2015,Barsanti:2016,Bayliss:2017}. 

The space density of clusters as a function of velocity dispersion 
also constrains the velocity bias in an assumed cosmology, and \citep{Rines:2007} find $b_v = 0.94 \pm 0.05$ and $1.28 \pm 0.06$ for WMAP1 and WMAP3 cosmologies, respectively. The quoted errors are statistical and based on a sample of 72 clusters in the SDSS DR4 spectroscopic footprint.  
The study of \citet{Maughan:2016} compares caustic masses derived from galaxy kinematics \citep[{\sl e.g.},][]{Diaferio:1999,Miller:2016} with X-ray hydrostatic masses. Such a comparison yields a measure of relative biases in hydrostatic and caustic methods, and their finding of $1.20^{+0.13}_{-0.11}$ for the ratio of hydrostatic to caustic $M_{500}$ estimates is consistent with unity at the $< 2\sigma$ level. If incomplete thermalization of the intracluster plasma leads hydrostatic masses to underestimate true masses by $20\%$ \citep[{\sl e.g.},][and references therein]{Rasia:2006}, then the central value of \citet{Maughan:2016} indicates that caustic masses would further underestimate true masses. Because of the relatively strong scaling $M \propto b_v^{-3}$, a value $b_v \simeq 0.9$ would suffice for consistency.  

Redshift space distortions provide another means to test velocity bias \citep{Tinker:2007}. The current constraints from  \citet{Guo:2015,GuoII:2015} indicate a magnitude-dependent bias, with $b^{-1}_{v}$ changing from slightly above one for bright systems --- the value $b_v = 1.05 \pm 0.08$ we employ in \S\ref{sec:massCalibration} to infer total mass --- to slightly below one for fainter galaxies.  Oddly, this trend is opposite to that inferred for galaxies from both hydrodynamic and N-body simulations, where bright galaxies are kinematically cooler than dimmer ones \citep{Old:2013, Wu:2013}.   The recent observational study of \citep{Bayliss:2017} finds a similar trend.

In summary, studies are in the very early stages of investigating velocity bias in the non-linear regime, both via simulations and in observational data. The statistical precision of future spectroscopic surveys, such as DESI \citep{DESI:2016}, will empower future analyses that may produce more concrete estimates of $b_v$ as a function of galaxy luminosity and host halo environment.  

Given the current level of systematic error in mass calibration, our ensemble velocity result is consistent with the weak-lensing mass calibration results of \citetalias{Lieu:2016}.
Similarly, the weak lensing results of \citet{Simet:2016} and \citet{Melchior:2017} for redMaPPer clusters agree with the  \citet{Farahi:2016} estimates.  
Better understanding of the relative biases of weak lensing, hydrostatic and other mass estimators will shed light on the magnitude of velocity bias in the galaxy population.


\section{Conclusion} \label{sec:conclusion}

We model ensemble kinetic motions of galaxies as a function of X-ray temperature to constrain a power-law scaling of mean galaxy velocity dispersion magnitude, $\langle \ln \sigmagal | \tthreeh, z \rangle$ for a sample of 132 spectroscopically confirmed C1 and C2 clusters in the XXL survey.  Spectroscopic galaxy catalogs derived from GAMA, SDSS DR10, VIPERS, VVDS and targeted follow-up surveys provide the input for the spectroscopic analysis.  From the kinetic energy, we derive total system mass using a precise dark matter virial calibration from N-body simulations coupled with a velocity bias degree of freedom for galaxies relative to dark matter. 

Following \citet{Rozo:2015} and \citet{Farahi:2016}, we employ a likelihood model for galaxy--cluster relative velocities, after removal of high-velocity outliers, and extract underlying parameters by maximizing the likelihood using an MCMC technique.  The analysis constrains the behavior of a primary Gaussian component, containing $\sim 90$\% of the non-outlier galaxies, the width of which scales as a power law with temperature, as anticipated by assuming self-similarity \citep{Kaiser:1986}. 

Based on 1908 galaxy-cluster pairs, we find a scaling steeper than self-similarity, 
\begin{equation}
\left\langle \ln \left( \frac{\sigmagal}{\kms} \right) | ~ \tthreeh, z=z_p \right\rangle  = \ln(\sigma_p) + \alpha \ln \left(\frac{\tthreeh}{2.2 \kev}\right)
\end{equation}
with $\sigma_p = 539\pm 16$ and $\alpha=0.63\pm 0.05$ at a pivot redshift of $z_p = 0.25$.  While redshift evolution is included in the likelihood, the data are not sufficiently dense at high redshift to establish a meaningful constraint on evolution.  

We identify and characterize several sources of systematic error and study the sensitivity of inferred parameters to the galaxy selection model and assumptions of the stacked model. The method is largely robust (Table~\ref{tab:sensitivityAnalysis}).  It is worth noting that these systematic error sources are generally different from those of other mass calibration methods, such as weak-lensing and hydrostatic X-ray methods, which allows the XXL survey to have an independent estimate of the cluster mass scale.

Employing the precise N-body virial mass relation calibrated in  \citet{Evrard:2008} coupled with an external constraint on galaxy velocity bias, $\sigmagal/\sigma_{\rm DM} = 1.05 \pm 0.08$, we derive a halo mass scaling
\begin{equation}
\begin{split}
& \left\langle \ln \left( \frac{E(z) M_{200}}{10^{14} \msol} \right) | ~ \tthreeh, z=z_p \right\rangle  = \\
  & ~~~~~~~~~~~~ \pi_{T} + \alpha_{T} \ln \left(\frac{\tthreeh}{2.2 \kev}\right) + \beta_{T} \ln \left(\frac{E(z)}{E(0.25)}\right)
\end{split}
\end{equation}
with normalization, $\pi_{T}=0.45 \pm 0.24$, and slopes, $\alpha_{T}=1.89 \pm 0.15$ and $\beta_{T}= -1.29 \pm 1.14$. 

Within the uncertainties, our result is consistent with mass scalings derived from both weak-lensing measurements of the XXL sample \citepalias{Lieu:2016} and provides an independent X-ray analysis using the hydrostatic assumption to obtain mass. 
But uncertainties in the scaling normalization remain at the level of $10-25\%$ (see Table~\ref{tab:comparisonObsScaling}), and fractional errors in slope are also of order ten percent.  

We note that the dominant source of uncertainty in our mass estimator is not statistical, but systematic uncertainty due to the galaxy velocity bias.  
Deeper and denser spectroscopic surveys, partnered with sophisticated sky simulations, will enable richer analyses than that performed here.  
As the accuracy of weak lensing and hydrostatic mass estimates improve, the ensemble method we employ here could be inverted to constrain the magnitude of velocity bias at small scales from future surveys such as DESI \citep{DESI:2016}. Such an approach has recently been applied to a small sample of Planck clusters by \citet{Amodeo:2017}. 

Larger numbers of spectroscopic galaxies at $z>0.5$ are needed to constrain the redshift evolution.  In recent hydrodynamic simulations that incorporate AGN feedback, 
\citet{Truong:2016} present evidence for weak redshift evolution in the slope of the mass-temperature scaling relation at $z < 1$, with stronger evolution at $z>1$.  
Next generation X-ray missions, such as eROSITA \citep{eROSITA:2012} and Lynx \citep{Surveyor:2015}, will offer the improved sensitivity needed to identify and characterize this population.  In the meantime, deeper XMM exposures over at least a subset of the XXL area can be used to improve upon the modest constraints on evolution we obtain using the current 10~ksec exposures.

The best practice in comparing the forthcoming, more sensitive observational data with theoretical models will require generating synthetic light-cone surveys from simulations and applying the same data reduction techniques to the models and observations.

An extension that we leave to future work is to properly include temperature errors into the ensemble spectroscopic likelihood model.  Richer data will allow investigation of potential modifications to the simple scaling model assumed here, including testing for deviations from self-similarity (in the redshift evolution of the normalization or a redshift-dependent slope, for example) and potential sensitivity to the assembly history or large-scale environment of clusters.

\begin{acknowledgements}
XXL is an international project based around an XMM-Newton Very Large Programme surveying two 25 deg$^2$ extragalactic fields at a depth of $ 5 \times 10^{-15}$~erg~cm$^{-2}$~s$^{-1}$ in the $[0.5-2]$ keV band for point-like sources. The XXL website is http://irfu.cea.fr/xxl. Multi-band information and spectroscopic follow-up of the X-ray sources are obtained through a number of survey programmes, summarised at http://xxlmultiwave.pbworks.com/. 
National Facility which is funded by the Australian Government for operation as a National Facility managed by CSIRO. GAMA is a joint European-Australasian project based around a spectroscopic campaign using the Anglo-Australian Telescope. The GAMA input catalog is based on data taken from the Sloan Digital Sky Survey and the UKIRT Infrared Deep Sky Survey. Complementary imaging of the GAMA regions is being obtained by a number of independent survey programmes including GALEX MIS, VST KiDS, VISTA VIKING, WISE, Herschel-ATLAS, GMRT and ASKAP providing UV to radio coverage. GAMA is funded by the STFC (UK), the ARC (Australia), the AAO, and the participating institutions. The GAMA website is http://www.gama-survey.org/. This paper uses data from the VIMOS Public Extragalactic Redshift Survey (VIPERS). VIPERS has been performed using the ESO Very Large Telescope, under the ``Large Programme'' 182.A-0886.
This work was supported by NASA Chandra grant CXC-17800360.
DR is supported by a NASA Postdoctoral Program Senior Fellowship at the NASA Ames Research Center, administered by the Universities Space Research Association under contract with NASA. 
FP acknowledges support by the German Aerospace Agency (DLR) with funds from the Ministry of Economy and Technology (BMWi) through grant 50 OR 1514 and grant 50 OR 1608.

\end{acknowledgements}

\bibliographystyle{aa}
\bibliography{scaling_rels}

\appendix

\section{Posterior parameter distributions} \label{app:posterior}

\begin{figure}
\centering
\includegraphics[width=0.95\linewidth]{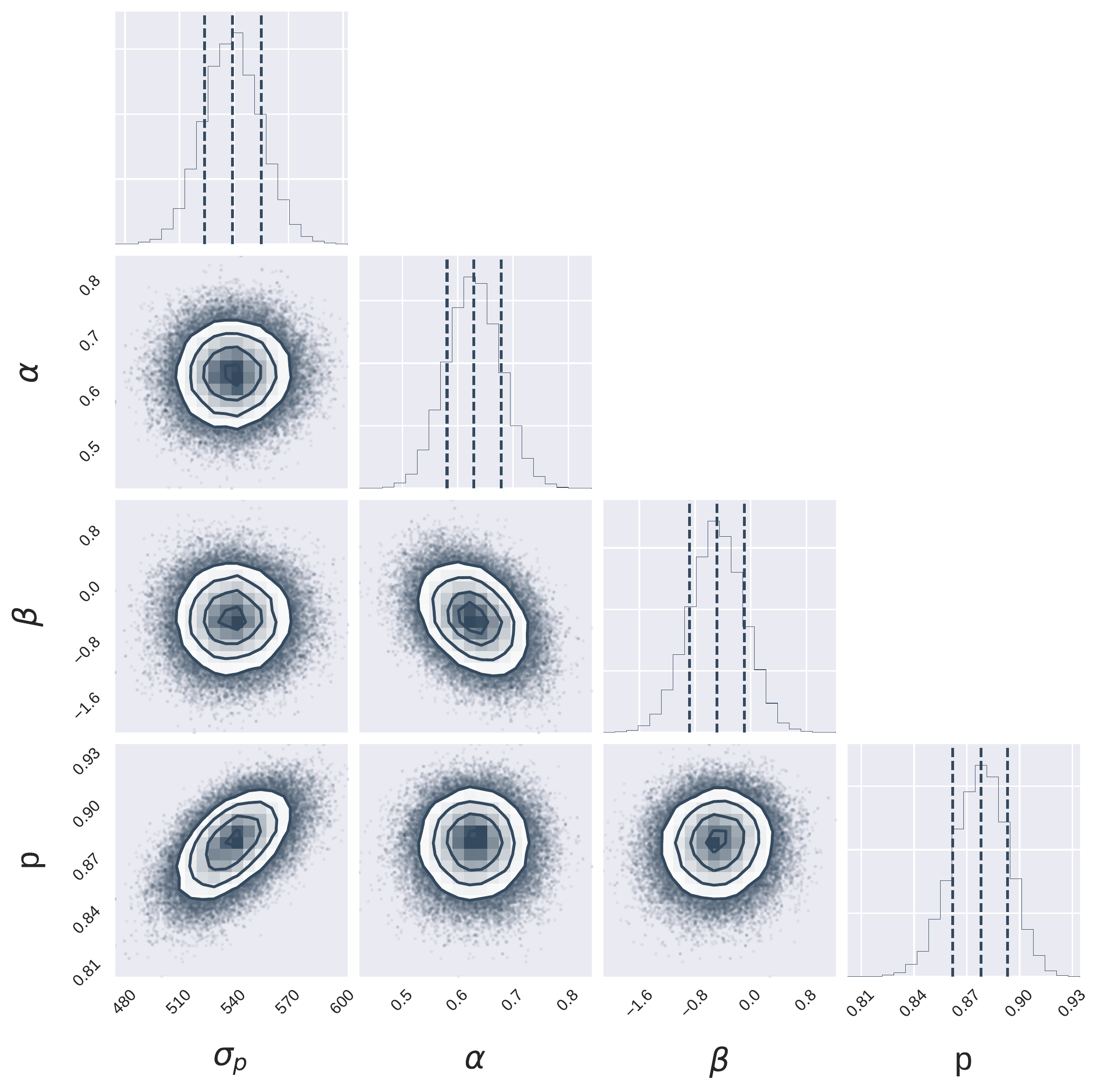}
\caption{Posterior likelihood distributions for $\sigmagal - \ktthreeh$ scaling relation parameters.}
\label{fig:constraints}
\end{figure}

Figure~\ref{fig:constraints} shows the posterior distributions of the free parameters of the fiducial model\footnote{Plot produced with the Python package \texttt{corner.py} \citep{corner}}. Posterior PDFs are close to Gaussian, illustrating the convergence of the MCMC chains. 

\section{Absolute r-band magnitude of selected galaxies} \label{app:r-band}

\begin{figure}
\centering
\includegraphics[width=0.90\linewidth]{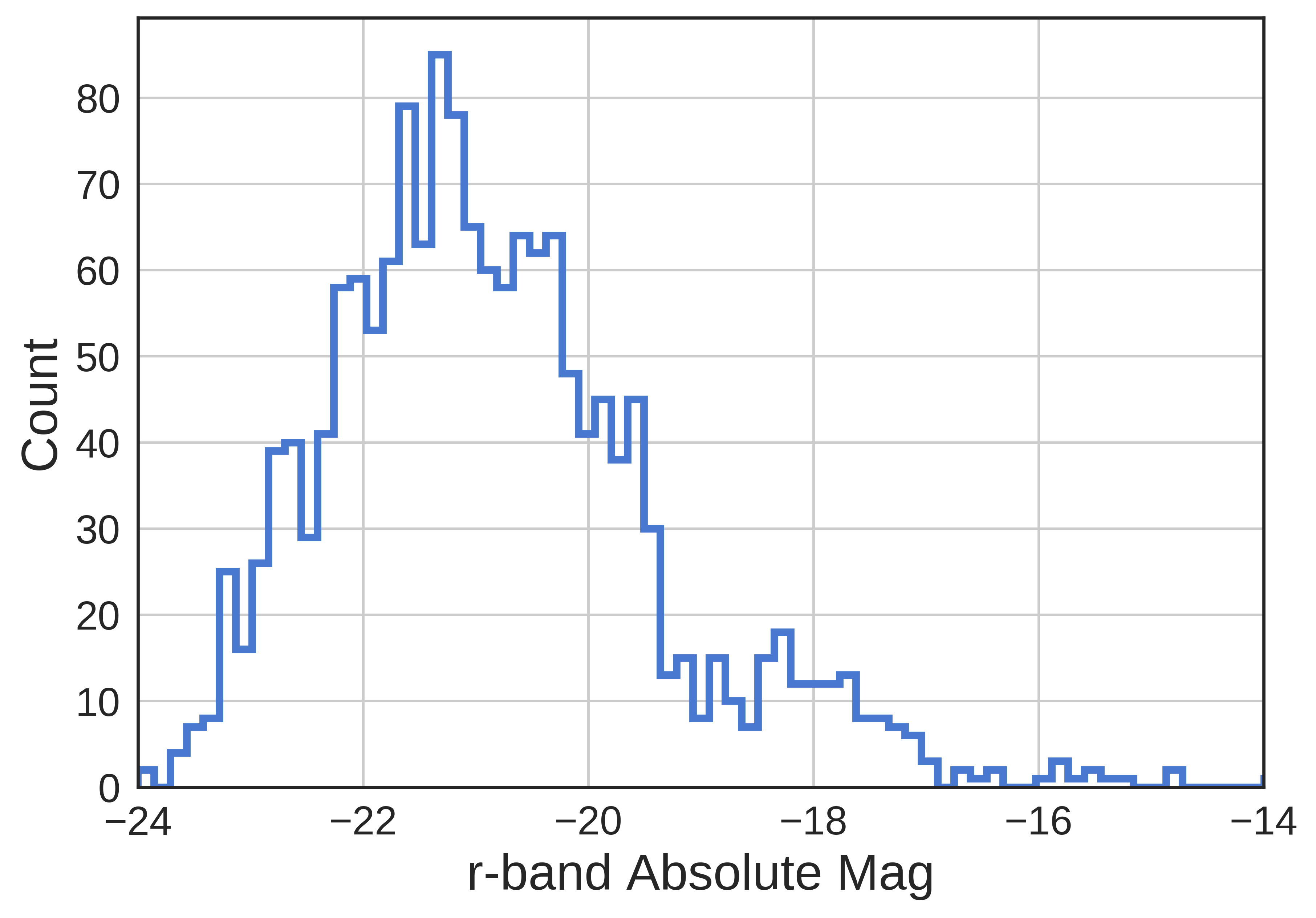}
\caption{The distribution of r-band absolute magnitude for selected galaxies after applying the fiducial aperture and velocity cuts.  
} 
\label{fig:r-band}
\end{figure}

 According to \citet{GuoII:2015} the velocity bias runs with the absolute magnitude of selected galaxies.  Figure \ref{fig:r-band} show the distribution of absolute r-band magnitude of selected galaxies in this work. We find that the peak of this distribution lies very near ${\rm M}_r = 21.5$, which justifies the choice of our prior distribution, $b_{v} = 1.05 \pm 0.08$ found by \citet{GuoII:2015} for this magnitude threshold.

\label{lastpage}
\end{document}